\documentclass[12pt]{article}

\usepackage{hyperref}
\hypersetup{
	hidelinks
}
\usepackage[cp1251]{inputenc}
\usepackage[english]{babel}

\usepackage{amsfonts,epsfig,epstopdf}
\usepackage{amssymb,amsmath,enumerate,float}
\usepackage[section]{placeins}
\makeatother

\addto\captionsenglish{}

\newcommand{\be}{\begin{equation}}
\newcommand{\ee}{\end{equation}}
\newcommand{\beq}{\begin{equation}}
\newcommand{\eeq}{\end{equation}}

\newtheorem{proposition}{Proposition}
\newtheorem{theorem}{Theorem}

\renewcommand{\b}{\mathbf{b}}
\renewcommand{\c}{\mathbf{c}}

\newcommand{\gR}{\mathbf{R}}
\newcommand{\gS}{\mathbf{S}}
\newcommand{\gK}{\mathbf{K}}

\title{Diffraction by a Dirichlet right angle on a discrete planar lattice}

\author{A. V. Shanin, A. I. Korolkov}

\begin{document}
\maketitle

\begin{abstract}

A problem of scattering by a Dirichlet right angle on a discrete square lattice is studied. 
The waves are governed by a discrete Helmholtz equation.
The solution is looked for in the form of the Sommerfeld integral. The Sommerfeld transformant 
of the field is built as an algebraic function.  The paper is 
a continuation of~\cite{Shanin2020}. 
  
\end{abstract}


\section*{NOTATIONS}
\begin{tabular}{cp{1\textwidth}}
  $\mathbb{C}$, $\overline{\mathbb{C}}$ & complex plane and Riemann sphere\\
  $K$ & wavenumber parameter of equation (\ref{eq201}) \\
  $u(m,n)$ & wave field on the lattice \\
  $\phi_{\rm in}$, $\phi$ & angle of propagation of the incident wave, angle of scattering \\
  $m$, $n$ & indexes of nodes in the discrete physical plane \\
  $x$, $y$ & wavenumber parameters \\ 
  $\gS_3$ & branched discrete plane, introduced in \cite{Shanin2020} \\
  $w_{m,n} (x,y)$ & plane wave (\ref{eq203}) \\
  $x_{\rm in}$, $y_{\rm in}$ & wavenumber parameters of the incident wave \\ 
  $\hat D(x, y)$ & dispersion function (\ref{eq205}) \\
  $y(x)$ & root of dispersion equation (\ref{eq207}) defined by (\ref{eq209})\\
  $\Upsilon (x)$ &  irrationality of $y(x)$,  (\ref{eq209b}) \\
  $\gR$ & Riemann surface of $y(x)$ or $\Upsilon (x)$ \\
  $\gR_3$ & a 3-sheet covering of $\gR$\\  
  $P_{3:1}$, $P_{1:0}$ & projections between $\gR_{3}$, $\gR_1$, and $\overline{\mathbb{C}}$ \\
  $\tilde X(x , j)$ , $\hat X(x , j)$ & notations for points of $\gR_3$ and $\gR$ \\  
  $\tilde x$, $\hat x$, $x$ & notations for points on $\gR_3$, $\gR$, and 
$\overline{\mathbb C}$ linked by natural projections \\  
  $\eta_{1,1}$, $\eta_{1,2}$, $\eta_{2,1}$, $\eta_{2,2}$ & branch points of $\gR$,
defined by (\ref{eq210}), (\ref{eq211})  \\
   $\Lambda$, $\Pi$, $\Pi'$ & 
     symmetries of $\gR_3$, see (\ref{eq:La}), (\ref{eq305}) \\  
  $A(\tilde x)$ & Sommerfeld transformant of the field (see representation (\ref{eq212})) \\
  $A_0$, $A_1$, $A_2$ & 
       the components of $A$ having different properties with respect to $\Lambda$ \\  
  $\tilde x_1$, $\tilde x_2$, $\tilde x_3$, $\tilde x_4$ & 
  prescribed poles of $A$ on $\gR_3$ \\    
  $Y_1, \dots , Y_4$ & residues of  $A$ \\ 
  $w_{m,n}(x,y)$ & discrete plane wave, (\ref{eq206}) \\
  $\Gamma_2$, $\Gamma_3$ & contours for the Sommerfeld integral \\	 
  $J_2$, $J_3$, $J_1'$, $J_3'$, $J_4'$ & contours encircling zero / infinity points on $\gR_3$\\ 
  $\gK_0$, $\gK_1$, $\gK_3$ & fields of functions meromorphic on $\overline{\mathbb C}$, 
   on $\gR$, and on $\gR_3$, respectively \\
  $\Omega_{j:l}$ & basis of extension $\gK_j$ over $\gK_l$ \\
  $F_1(x)$, $F_2(x)$ & nontrivial elements of the basis $\Omega_{3:1}$ \\ 
  $\varpi$ & $\exp\{2 \pi i / 3\}$, cubic root of 1 \\
  $\hat \b$, $\b$  & an important point on $\gR$ used for building $F_1$ and its affix \\ 
  $\chi(\hat x)$, $T_\alpha$, $T_\beta$ & Abelian integral of the first kind on $\gR$ 
  (see (\ref{eq005}))   and its periods \\
  $\alpha$, $\beta$ & natural coordinates on the torus $\gR$\\      
 $\psi$ & mapping $\chi \to \hat x$
 					 
\end{tabular}

\section{Introduction}
This paper continues the research presented in~\cite{Shanin2020}.
A 2D discrete square lattice is under consideration. The lattice bears a discrete 
Helmholtz equation with a 5-point stencil. The first quadrant of the lattice is  
blocked by setting the field equal to zero there. The problem of diffraction of 
an incident plane wave by the blocked angle is studied. The motivation and the literature review for such a problem can be found in \cite{Shanin2020}.  

A new formalism has been developed for this problem.
Similarly to continuous problems of diffraction in angular domains, a branching 
surface $\gS_3$ 
is introduced in the physical discrete plane, and the diffraction problem is reformulated 
as a propagation problem on this surface using the reflection principle.
An analog of the Sommerfeld integral for field representation 
is introduced.
This integral is a contour integral on the dispersion diagram of the lattice, which is a compact  
Riemann surface $\gR$ equipped with a structure of a 
complex manifold. Topologically, $\gR$ is a torus. 
The integrand is a differential form that is multivalued on the 
dispersion diagram and possesses prescribed 
poles corresponding to the incident wave and reflected waves. 
The contour of integration depends on the position of the observation point, and ``slides'' along the surface as the  observation 
point moves. We introduce a 3-sheet covering of $\gR$, named $\gR_3$, to take into account 
the multivaluedness of the integrand.

The integrand form contains an unknown function 
referred to as the Sommerfeld transformant of the field.
It obeys a certain functional problem. 
In \cite{Shanin2020} the authors 
found this transformant in terms of elliptic functions. However, 
such a representation is not convenient. Moreover, it can be  proven that        
such a transformant should be an algebraic function, thus, a representation 
through the elliptic functions is somewhat unnecessarily complicated. The aim of the 
current paper is to build the Sommerfeld transformant of the field as an algebraic function,
and then to study the properties of the field. 

We should note that Riemann surfaces and Abelian integrals have been used for solving 
diffraction problem before. The context of the application of this theory was the 
Riemann--Hilbert problem on Riemann surfaces \cite{Zverovich1969,Zverovich1971,Antipov1991,Antipov2002}. 

The paper is organized as follows. 

In {\bf Section~\ref{sec2}} some preliminary steps are made. 
The initial diffraction problem is formulated and is reformulated as a 
propagation problem on a branched lattice~$\gS_3$.  
The Riemann surface $\gR$ related to the dispersion diagram of the lattice is introduced. 
The Riemann surface $\gR_3$ describing all plane waves on $\gS_3$ is built. Both 
Riemann surfaces are equipped with a structure of a complex manifold. 
The Sommerfeld integral for the field is written, and Functional problem~1 for the 
Sommerfeld transformant $A$ is formulated. 
The functional problem is constituted in finding a function meromorphic on a given Riemann
surface, having prescribed poles on it with known residues.  

In {\bf Section~\ref{sec_math_bas}}   
we introduce some basic mathematical concepts that are necessary to solve the 
functional problem. We introduce the transformations of $\gR_3$ and $\gR$ (the desk transformations 
in terms of \cite{Khovanskii2014}). These transformations enable to convert Functional problem~1
into a set of partial problems that are slightly simpler. 
The key idea for solving the functional problem is to introduce for each Riemann surface 
a field (in the algebraic meaning of this word) of functions meromorphic on this surface. The Sommerfeld transformant 
belongs to some of these fields, $\gK_3$, which is an extension of a simpler field $\gK_1$. 
Then, one can construct a basis of this extension, which is a set of three functions $[1, F_1 , F_2]$, 
two of which are quite difficult to find. Once the basis is built, 
one can find the function $A$ by constructing relatively simple coefficients of the expansion. 

In {\bf Section~\ref{sec_basis}} we build the basis function $[1, F_1 , F_2]$. 
This is a tricky part of the paper. The functions $F_{1,2}$ are found 
as cubic roots of functions $G_{1,2}$ specially tailored on the base of Abel's theorem. 
The Abel's theorem is then replaced by an algebraic condition.

In {\bf Section~\ref{sec_span}} we build the coefficients of expansion of the Sommerfeld transformant
$A$ using the basis $[1, F_1 , F_2]$. This is done in an elementary way by studying the zeros and poles of 
$F_1$, $F_2$, and $A$.

In {\bf Section~\ref{sec_num}} a detailed description of the numerical procedure is given, and some results are demonstrated. 

The {\bf Appendix} contains the proof of the validity of the Sommerfeld representation of the wave field, 
some statements completing the algebraization of Abel's theorem, and the proof of the theorem on which the computation of the 
functions $F_{1,2}$ is based.

The paper is written mainly in the Theorem / Proof / Remark style for simplicity of understanding.  
   
There are two important issues, clearly related to the diffraction problem considered in the current paper, namely, 
they are the computation of the directivity of the far field and building a low-frequency approximation (the limiting case of the
continuous medium). Both issues are interesting and not elementary. We do not include them into the paper due to the lack of space, 
and are planning to write a separate work on them.   


\section{The Sommerfeld integral for a discrete wedge diffraction problem}
\label{sec2}
\subsection{Problem formulation}
\label{wedge1}

\begin{figure}[ht]
\centering
\includegraphics[width=0.4\textwidth]{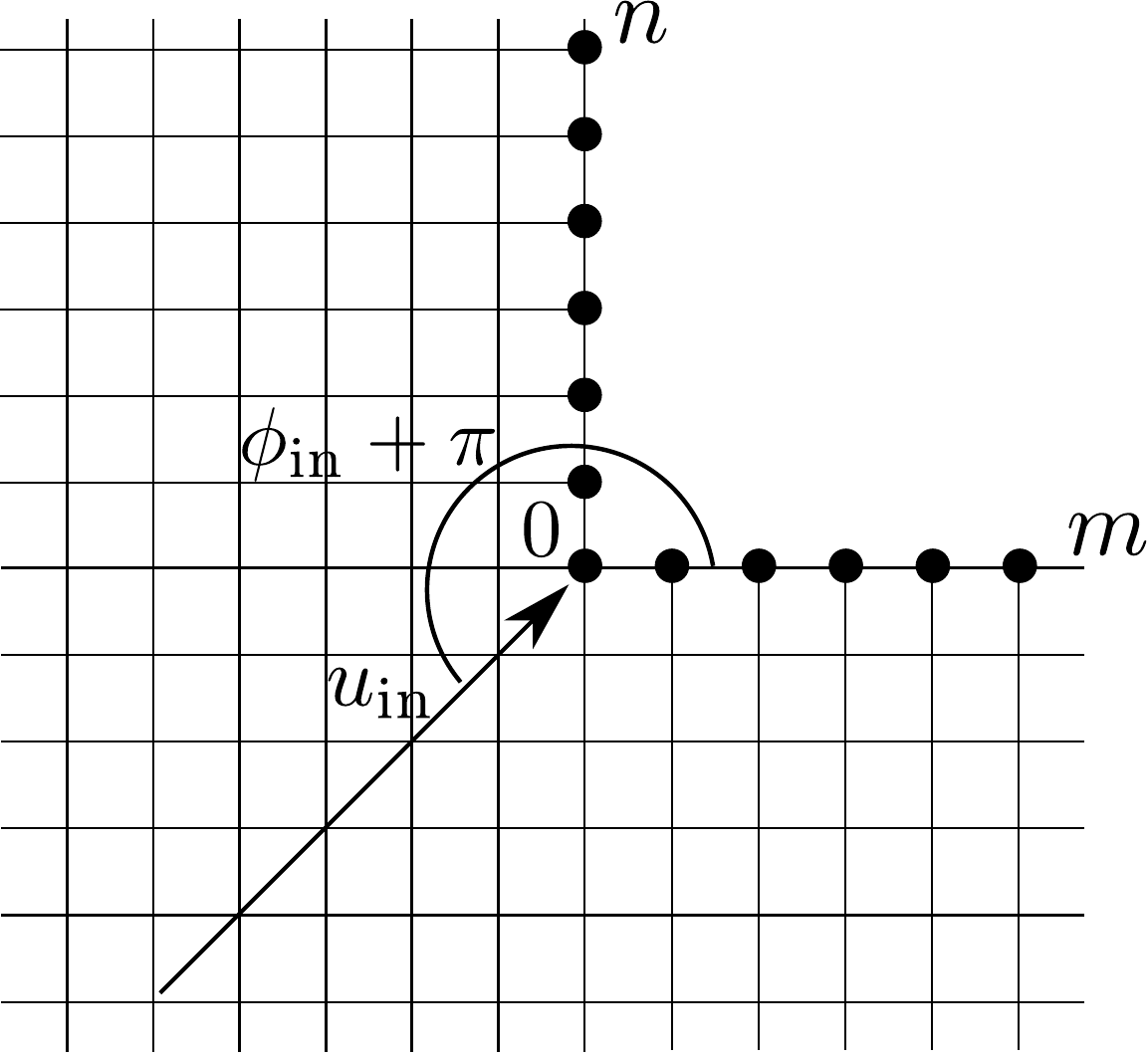}
\caption{Geometry of the problem of diffraction by an angle. Black circles show the position of the Dirichlet boundary (blocked nodes)}
\label{fig12b}
\end{figure}

Consider a planar square lattice whose nodes have integer indices $(m,n)$. 
Let the homogeneous discrete Helmholtz equation
\begin{equation}
u(m,n-1) + u(m, n+1) + u(m-1, n) + u(m+1 , n) + (K^2 - 4) u(m,n) = 0 
\label{eq201}
\end{equation}
be valid in the domain
\[
m < 0 \quad \mbox{or} \quad  n < 0
\] (see Fig.~\ref{fig12b}). 
The wavenumber parameter $K$ has a positive real part and a small positive
imaginary part corresponding to an energy absorption. 

The set of nodes with 
\[
(m = 0\mbox{ and }n \ge 0) 
\mbox{ or } 
(n = 0\mbox{ and }m \ge 0) 
\]
is the boundary of the domain. We assume that this boundary is of the 
Dirichlet type, so 
\begin{equation}
u(m,n)  = 0
\label{eq202}
\end{equation} 
on it. 

In order to describe the incident wave, 
we introduce 
plane waves on such a lattice:
\begin{equation}
w_{m,n} = w_{m,n} (x , y) = x^m y^n,
\label{eq206}
\end{equation}
provided that the pair of wavenumber parameters $(x , y)$
obey the dispersion equation
\begin{equation}
\hat D (x , y) = 0,
\label{eq207}
\end{equation}
\begin{equation}
\hat D(x , y) \equiv x + x^{-1} + y + y^{-1} + K^2 - 4.
\label{eq205}
\end{equation}
One can see that (\ref{eq207}) guarantees fulfillment of 
the homogeneous Helmholtz equation (\ref{eq201})  by~$w$. 

The total wave for our problem is a sum of the incident wave  and 
the scattered wave: 
\[
u(m,n) = u_{\rm in} (m,n) +  u_{\rm sc} (m,n) , 
\]
where   
\begin{equation}
u_{\rm in} (m,n) =w_{m,n} (x_{\rm in} , y_{\rm in}) = x^m_{\rm in} y^n_{\rm in},
\label{eq203}
\end{equation}
is the incident plane wave,
and $x_{\rm in}$ and $y_{\rm in}$ are wavenumber parameters. Indeed, they 
obey the dispersion equation:
\begin{equation}
\hat D(x_{\rm in} , y_{\rm in}) = 0,
\label{eq204}
\end{equation}

For simplicity of the problem formulation, we assume that the wave travels into the direction of positive 
$m$ and $n$. Since the waves have some attenuation, this means that 
\[
|x_{\rm in}| < 1 ,\qquad |y_{\rm in}| < 1.
\]
We introduce the angle of incidence 
by the relation 
\begin{equation}
\phi_{\rm in} \equiv \arctan
\left( 
\frac{
y_{\rm in} - y^{-1}_{\rm in}
}{
x_{\rm in} - x^{-1}_{\rm in}
}
\right).
\label{eq_phiin}
\end{equation}
Such a definition of the angle may seem not obvious, but it is  motivated by the
saddle point argument in \cite{Shanin2020}, see equation (47) there. 
We assume that angle $\phi_{\rm in}$ is real, and 
\begin{equation}
0 < \phi_{\rm in} < \pi / 2.
\label{eq_inc}
\end{equation} 

The scattered wave $u_{\rm sc}$ should obey the radiation condition, i.~e.\ it should decay 
at infinity.  The aim is to find $u_{\rm sc}$.

\vskip 6pt
\noindent
{\bf Remark.}
For a $K$ with ${\rm Im}[K^2] \ne 0$ and under the condition (\ref{eq_inc})
the scattered field  $u_{\rm sc}$ belongs to the space $l_2(\mathbb{Z}^2)$. The 
solution is unique. The proof of uniqueness can be found in Appendix~4.


\subsection{Reformulation of the diffraction problem on a branched discrete lattice}
\label{ss_21}

In the current paper we use the Sommerfeld integral technique for an angular domain of a 
discrete lattice. It is well-known that the Sommerfeld method starts with applying the principle of 
reflection,  with the aim to get rid of the scatterers and to obtain a diffraction problem on a branched surface. 

Introduce the angle 
\begin{equation}
\tan \phi = n/m,
\label{eq_an}
\end{equation}
such that the domain shown in Fig.~\ref{fig12b} is 
$\pi / 2 \le \phi \le 2\pi$.
Take the solution $u(m,n)$ in this domain. 
Reflect the domain with respect to the horizontal axis and define the function  $-u(m,-n)$ on it.
Connect the initial domain and the reflected domain by merging the nodes  on the boundaries at $m \ge 0$, $n = 0$. 
As the result, get the discrete angular domain $\pi /2 \le \phi \le 7\pi / 2$
with some function $u(m,n)$ defined on it. 
The following proposition is valid: 

\begin{proposition}
The Helmholtz equation (\ref{eq201})
is valid on the merged nodes $m > 0$, $n = 0$.
\end{proposition}

To prove this proposition, one can directly check (\ref{eq201}) at corresponding nodes. The value of $u (m,n)$ 
is zero at them due to the boundary condition, and $u(m,n)$ is odd with respect to $n$ by construction.  
 
\begin{figure}[ht]
\centering
\includegraphics[width=0.7\textwidth]{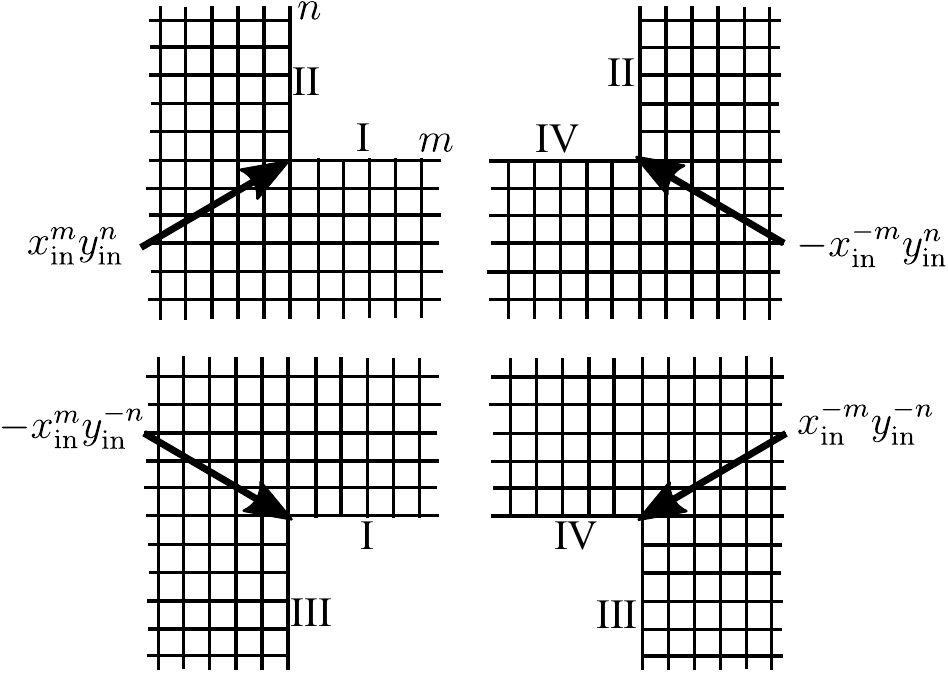}
\caption{Riemann surface $\gR$}
\label{figS3}
\end{figure}

Thus, we applied the reflection principle once, deleted a horizontal part of the boundary, and obtained a wider discrete angular domain.  By repeating this procedure two more times, now with respect to the vertical part of the boundary, obtain a function $u$ defined on a branched discrete surface 
$\gS_3$ shown in 
Fig.~\ref{figS3}. 
The boundaries that should be merged with each other are shown by equal Roman numbers. 

The surface $\gS_3$ has three points over each point $(m,n) \ne (0,0)$; 
the  origin is the branch point, thus
it has three sheets over a discrete planar lattice,
and 
any single-valued function on $\gS_3$ is periodic with respect to $\phi$ with the period 
$6\pi$, i.~e.\ the branch point has order three.
The function $u(m,n)$ built on $\gS_3$ by the reflections obeys  
(\ref{eq201}) at any point of $\gS_3$ except the origin.
The equation connects $u$ at some node with its four neighbors in the mesh.

There are four incident plane waves on $\gS_3$ (shown in Fig.~\ref{figS3}). 
One can summarize these waves in the following table:

\vskip 6pt
\begin{center}
\begin{tabular}{c|c|c}
angle of incidence $\phi'$ & amplitude &  formula \\
\hline
$\phi_{\rm in} + \pi$ & $1$  & $x^m_{\rm in} y^n_{\rm in}$ \\
$3\pi - \phi_{\rm in}$  & $-1$  & $-x_{\rm in}^m y^{-n}_{\rm in}$ \\
$\phi_{\rm in} + 4\pi$ & $1$  & $x^{-m}_{\rm in} y^{-n}_{\rm in}$ \\
$6\pi - \phi_{\rm in}$   & $-1$  & $-x_{\rm in}^{-m} y_{\rm in}^n$ \\
\end{tabular}
\end{center}
\vskip 6pt

Each plane wave is visible in the domain $\phi' - \pi < \phi < \phi' + \pi$, 
where $\phi'$ is the angle of incidence of the corresponding wave. 

As it is shown in \cite{Shanin2020}, the Sommerfeld integral provides a solution 
of the diffraction problem on $\gS_3$, i.~e.\ 
it yields a function $u(m,n)$ single valued on $\gS_3$, obeying 
(\ref{eq201}) everywhere except the origin, and composed of the incident waves 
(each at its visibility domain), and the scattered field decaying as $\sqrt{m^2 + n^2} \to \infty$.  
The restriction of such a solution onto the angle $\pi / 2 \le \phi \le 2\pi$ is 
a solution of the initial diffraction problem. 

\vskip 6pt
\noindent
{\bf Remark.}
The structure of the problem on $\gS_3$ gives important clues to the functional problem for the 
Sommerfeld's transformant. Namely, it makes clear why it is necessary to study a 3-sheet covering of 
$\gR$ (see below), and why the transformant has four poles. For the first question, as it is shown in 
\cite{Shanin2020}, the number of sheets of the branched discrete lattice is equal to the 
multiplicity of the transformant over~$\gR$. For the second question, each pole corresponds 
to a plane incident wave on $\gS_3$.


\subsection{Riemann surfaces  $\gR$ and $\gR_3$}
\label{wedge2a}

\begin{figure}[ht]
\centering
\includegraphics[width=0.8\textwidth]{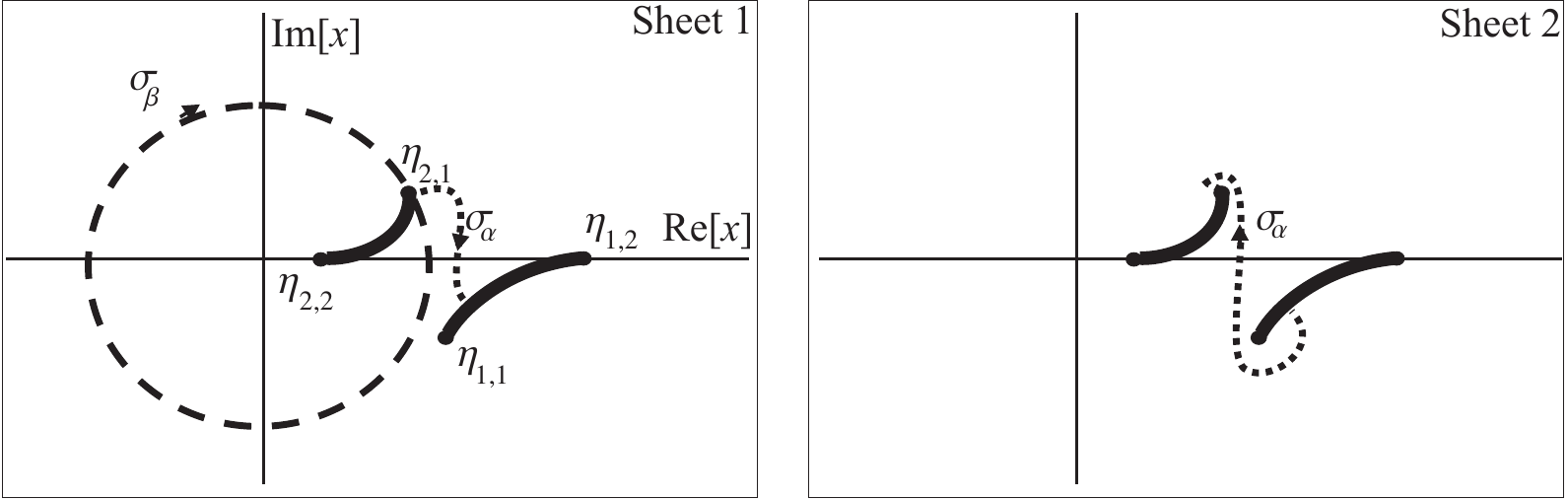}
\caption{Contours $\sigma_{\alpha}$ and $\sigma_{\beta}$ on $\gR$}
\label{fig22}
\end{figure}

The equation (\ref{eq207}) can be solved with respect to 
$y$ for some fixed $x$:
\begin{equation}
y(x)  = y_{\pm} (x) = - \frac{K^2 - 4 + x + x^{-1}}{2} \pm 
\frac{ \sqrt{(K^2 - 4 + x + x^{-1})^2-4}}{2}
\label{eq208x}
\end{equation}
The function $y(x)$ is a two-valued function.
It is easy to check that its two values  have the property 
$y_+ (x) y_- (x) =1$.
The Riemann 
surface  of $y(x)$ is denoted by $\gR$ and is shown in 
Fig.~\ref{fig22}.

Branch points of $\gR$ are 
$\eta_{1,1}$, $\eta_{1,2}$, $\eta_{2,1}$, $\eta_{2,2}$:
\begin{equation}
\eta_{1,1} = -\frac{d}{2} - \frac{i \sqrt{4 - d^2 }}{2} ,
\quad 
\eta_{2,1} = -\frac{d}{2} + \frac{i \sqrt{4 - d^2 }}{2} ,
\quad 
d = K^2 - 2, 
\label{eq210}
\end{equation} 
\begin{equation}
\eta_{1,2} = -\frac{d}{2} + \frac{\sqrt{d^2 - 4}}{2} ,
\quad 
\eta_{2,2} = -\frac{d}{2} - \frac{\sqrt{d^2 - 4}}{2} ,
\quad 
d = K^2 - 6 .  
\label{eq211}
\end{equation} 
All branch points are of the second order. One can see that 
\[
y(\eta_{2,1})= y(\eta_{1,1}) =1 , \qquad  y(\eta_{2,2}) = y(\eta_{1,2}) = -1.
\]
The branch points 
are connected by cuts on $\gR$ (shown by bold lines in the figure). 
The cuts are conducted in such a way that 
$|y(x)| = 1$ on them.
This corresponds to the conditions 
\begin{equation}
{\rm Im}[K^2 - 4 + x + x^{-1}] = 0, 
\qquad
-2<{\rm Re}[K^2 - 4 + x + x^{-1}] <2. 
\label{eq211x}
\end{equation}
The sides of the cuts 
marked by the same Roman numbers are attached to each other.  

We select the {\em physical sheet\/} (or sheet~1) of $\gR$ as the sheet on which 
$|y| \le 1$. 
Let $\Xi (x)$ be the value of the function $y(x)$ on the physical sheet, i.~e.\
\begin{equation}
\Xi(x) = - \frac{K^2 - 4 + x + x^{-1}}{2} + 
\frac{ \sqrt{(K^2 - 4 + x + x^{-1})^2-4} }{2},
\label{eq209}
\end{equation}
where the square root is taken in its ``arithmetical'' sense (with a cut along the negative 
half-axis of the argument). 
Obviously,
\begin{equation}
\Xi^{-1}(x) = - \frac{K^2 - 4 + x + x^{-1}}{2} - 
\frac{ \sqrt{(K^2 - 4 + x + x^{-1})^2- 4 }}{2},
\label{eq209a}
\end{equation}
and this is the value of $y(x)$ on sheet~2. 

The Riemann surface $\gR$ is compactified, i.~e.\ 
the infinite points of sheet~1 and of sheet~2 are added
(this means that $x$ takes values on the Riemann sphere~$\overline{\mathbb{C}}$).
One can check that $\Xi (\infty)  = 0$. 
It is important that the infinite points are not branch points of~$\gR$,
i.~e.\ a bypass about an infinity does not change the sheet of the Riemann 
surface.  

The authors made some efforts in \cite{Shanin2020} to demonstrate
that $\gR$ is topologically {\rm a torus}. In particular, Fig.~4 and 
Fig.~6 of 
\cite{Shanin2020} demonstrate coordinates $(\alpha, \beta)$ on the torus, 
both taking values on a circle $0 \le \alpha, \beta < 2\pi$.

Introduce the function $\Upsilon(x)$ single-valued on $\gR$ 
and defined on the physical sheet (sheet~1) by  
\begin{equation}
\Upsilon(x) = x(\Xi(x) - \Xi^{-1}(x)) = 
x \sqrt{(K^2 - 4 + x + x^{-1})^2 - 4}.
\label{eq209b}
\end{equation}
The square root takes the arithmetical value. 
Indeed, on sheet~2 this function is equal to 
\[
\Upsilon(x) = - x \sqrt{(K^2 - 4 + x + x^{-1})^2 - 4}.
\]
The importance of this function is explained by its equivalent form: 
\begin{equation}
\Upsilon(x) = 
\sqrt{(x  - \eta_{1,1})(x  - \eta_{1,2})
(x  - \eta_{2,1})(x  - \eta_{2,2})} .
\label{eq209ba}
\end{equation}

One can see that on the physical sheet
\begin{equation}
y(x) = - \frac{K^2 - 4 + x + x^{-1}}{2} + \frac{\Upsilon (x)}{2 x} .
\label{useful}
\end{equation}
The function $\Upsilon (x)$ is the irrationality of $\Xi(x)$, 
thus
the Riemann surface of $\Upsilon(x)$ is the same as of $y (x)$, i.~e.\ this 
surface is~$\gR$.

\begin{figure}[ht]
\centering
\includegraphics[width=0.8\textwidth]{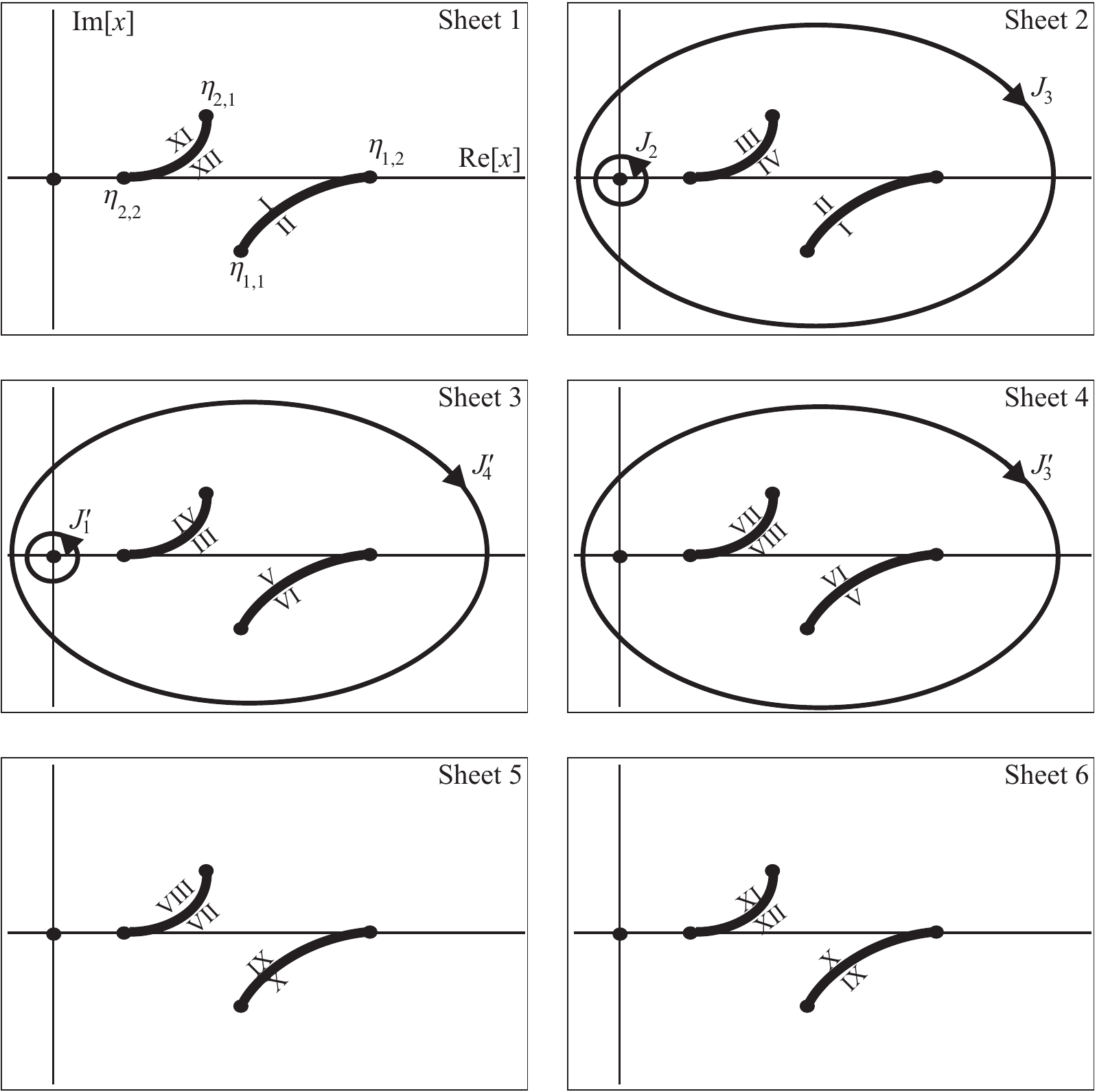}
\caption{Riemann surface $\gR_3$ and integration contours on it}
\label{fig03}
\end{figure}

Similarly to \cite{Shanin2020},
construct Riemann surface~$\gR_3$ as follows.
Take six  copies of the compactified complex plane of $x$ (the sheets), make cuts in them (the same as for $\gR$, 
see by (\ref{eq211x})), 
and assembly surface according to scheme shown in Fig.~\ref{fig03} by Roman numbers.
Note that this surface is {\em not\/} introduced as a Riemann surface of a certain function,
and our aim below is to find such functions.

Introduce notations for the points of the compactified complex plane $\mathbb{\bar C}$ 
and for the Riemann surfaces $\gR$,~$\gR_3$. 
The points of $\gR_3$ will be indicated by the $\tilde \cdot$ decoration, 
the points of $\gR$ will be indicated by the $\hat \cdot$ decoration,
and the  points of $\mathbb{\bar C}$ will exist without decorations. For example, 
\[
\tilde x \in \gR_3, 
\qquad 
\hat   x \in \gR,
\qquad  
x \in \mathbb{\bar C}.
\]

Introduce the projections
\[
\tilde x \stackrel{P_{3:1}}{\longrightarrow} \hat x \stackrel{P_{1:0}}{\longrightarrow} x .
\]
Let both projections keep the {\em affix\/} (the value of $x$), so, the 
projections $P_{1:0}(\cdot)$ and $ P_{1:0}(P_{3:1}(\cdot))$   take an 
affix of a point of a Riemann surface. 
Let the projection $P_{3:1}$ map the sheets 1, 3, 5 of $\gR_3$
onto  sheet~1 of $\gR$, and the sheets 2, 4, 6 of $\gR_3$
onto  sheet~2 of $\gR$. The points on the cuts are served by continuity 
and cause no problem. 

We keep the following convention in the whole paper: 
everywhere $\hat x$ is the projection of $\tilde x$, and $x$ is a projection of $\hat x$
and $\tilde x$. Indeed, this is valid for any letter instead of $x$ (it may be, say,
$a$ or $\b$). This can be written as 
\begin{equation}
\hat \cdot \equiv P_{3:1} (\tilde \cdot),
\qquad 
\cdot \equiv P_{1:0} (\hat \cdot) \equiv P_{1:0} (P_{3:1} (\tilde \cdot)), 
\label{eq_proj}
\end{equation}
where $\cdot$ stays for any letter, possibly with indexes, but without a decoration.

To specify the points $\tilde x$ on $\gR_3$ and $\hat x$ on $\gR$, we introduce notation
$\tilde X(x, j)$ and $\hat X (x, j)$. 
The notation $\tilde x = \tilde X(x, j)$ denotes the point having the affix $x \in \overline{\mathbb{C}}$ 
and lying on the sheet number $j \in \{ 1,2,3,4,5,6 \}$, 
as it is shown in Fig.~\ref{fig03}. 
Similarly, for the notation $\hat x = \hat X (x, j)$ the index $j$ takes values in $\{ 1 , 2\}$, and the 
sheets are shown in  Fig.~\ref{fig22}.  
Indeed, 
\[
P_{1:0} ( \hat X (x , j) ) = x, 
\]
\[
P_{3:1} ( \tilde X (x , j) ) = \hat X (x , j'),
\qquad 
j' = 1 \mbox{ for } j = 1,3,5,
\quad
j' = 2 \mbox{ for } j = 2,4,6.
\]

One can check 
directly that:

\begin{proposition}
\label{prop_cont1} 
Projections $P_{3:1}$ and $P_{1:0}$ are continuous. 
\end{proposition}

This fact is important. Namely,  
$\gR_3$ is a 3-sheet {\em covering\/} of $\gR$ (see \cite{Donaldson2011}).
Note that 
each point  $\hat x \in \gR$ (including the branch points) has exactly three preimages $P_{3:1}^{-1} (\hat x)$. 
A covering of a torus without branch points is also a torus, thus 
$\gR_3$ has topology of a torus. 
We explored this feature in \cite{Shanin2020} (see Fig.~13 there).



If function $f(\hat x)$ is single valued on $\gR$, there is no difficulty to define it on 
$\gR_3$ as a single-valued function 
\[
f(\tilde x) = f(P_{3:1} (\tilde x)) = P(\hat x).
\] 
Conversely, if $f (\tilde x)$ is single-valued on $\gR_3$, the function 
$f(\hat x)$ is generally three-valued on~$\gR$, and $f(x)$ is six-valued 
on $\mathbb{\bar C}$.
Functions $\Upsilon (\hat x)$ and $y(\hat x)$ are single-valued; 
functions
$\Upsilon (x)$ are $y(x)$ are two-valued.


Consider the oriented contours $\sigma_\alpha$ and $\sigma_\beta$ on $\gR$ 
(see Fig.~\ref{fig22}) starting and ending at $\eta_{2,1}$.  
Contours $\sigma_\alpha$ and $\sigma_\beta$ play an important role in 
\cite{Shanin2020} and here.
The contour $\sigma_\alpha$ is homotopic to the ``real waves''
contour on~$\gR$, i.~e.\ to the contour on which the propagation 
angles 
\[
\phi = \arctan \left( \frac{y(x) - y^{-1}(x)}{x - x^{-1}}  \right)
\]
are real, and which tends to an arc of the unit circle as ${\rm Im}[K] \to 0$. 
The points on the ``real waves'' contour become saddle points when the far field is estimated~\cite{Shanin2020}.
The contour $\sigma_\beta$ is homotopic to the unit circle 
on the physical sheet of $\gR$;  
this is the integral path for the Green's function of 
a discrete plane (see \cite{Shanin2020}).  
The contours $\sigma_\alpha$ and $\sigma_\beta$ form a {\em canonical dissection\/}
of $\gR$, i.~e.\ $\gR$ cut along $\sigma_\alpha$ and $\sigma_\beta$ becomes simply connected.


The paths $\sigma_\alpha$ and $\sigma_\beta$ are generators of the fundamental group 
$\pi_1$ of $\gR$ (note that $\pi_1$ is commutative). Thus, many topological properties of 
$\gR$ are connected with $\sigma_\alpha$ and $\sigma_\beta$. 
In particular, 
comparing Fig.~\ref{fig22} with Fig.~\ref{fig03}, we find that 

\begin{proposition}
\label{prop_cont}

a) The preimage $P^{-1}_{3:1} (\sigma_\beta)$ 
is a set of three copies of $\sigma_\beta$ (on sheets~1,3,5).
The preimage $P^{-1}_{3:1} (\sigma_\alpha)$
is a connected three-sheet covering of $\sigma_\alpha$. 

b) Let $\gR'$ be any covering of $\gR$, and let 
$P: \, \gR' \to \gR$ be the corresponding projection. Let $P^{-1} (\sigma_\beta)$
be three copies of $\sigma_\beta$, and let $P^{-1} (\sigma_\alpha)$
be a connected three-sheet covering of $\sigma_\alpha$. Then
$\gR'$ is equivalent to~$\gR_3$.   
\end{proposition}


\subsection{A structure of a complex manifold  on $\gR_3$ and contour integration}   

Introduce a structure of a
complex manifold on $\gR$ and on~$\gR_3$. 
By definition~\cite{Shabat1992},
a complex manifold is a 
union of possibly intersecting neighborhoods $\mathcal{U}_s$ in each of which 
a local complex variable $\tau_s$ can be introduced,
describing the neighborhood in a trivial 
way (there exists a continuous bijection between $\mathcal{U}_s$ and some 
open circle in the $\tau_s$ plane). 
Transitions between the local variables in the intersections of the neighborhoods 
should be biholomorphic. 

For both $\gR$ and $\gR_3$, 
a possible choice of local variables is as follows.  
In each small neighborhood not including the infinity or $\eta_{j,l}$ (i.~e.\ almost everywhere), 
one can take 
$\tau_s = x$ as a local variable. 
In the neighborhoods of the points with $x = \eta_{j,l}$
one can choose 
$\tau_s = \sqrt{x - \eta_{j,l}}$. 
The presence of the square root provides a one-to-one correspondence between a
neigborhood of zero in the domain of the local variable and the neigborhood of the branch point 
of $\gR$ or~$\gR_3$.   
Near the infinities one can take $\tau_s = x^{-1}$.
%

Indeed, one can introduce the structure of a complex manifold on the whole  
$\overline{\mathbb{C}}$ by taking the local variable $\tau_s = x^{-1}$ at the neighborhood of the infinity 
and $x$ in the finite part of~$\mathbb{C}$.

Thus, $\overline{\mathbb{C}}$, $\gR$, and $\gR_3$ become complex manifolds, and one can make an important note: 
all points of $\overline{\mathbb{C}}$, 
$\gR$, or $\gR_3$ (including the infinities and the branch points) are regular from the point of view of the 
complex structure.

Define a single-valued  function $f$ on $\gR$ or on $\gR_3$.   
At each 
neighborhood $\mathcal{U}_s$ one can express it as a function of the local variable:
$f = f(\tau_s)$. The function $f$ is analytic in $\mathcal{U}_s$ if $f(\tau_s)$ 
is analytic. Thus, definition of analyticity becomes local. 
A function has a pole or zero at some point belonging to $\mathcal{U}_s$
if $f(\tau_s)$ has a pole or zero at the corresponding point. 
The order of the pole/zero of a function is also defined with respect to the local variable.  

The concept of a pole / zero at some point  
on the complex manifolds 
$\gR$ or on $\gR_3$ differs from that on $\overline{\mathbb{C}}$.
For example the function $\Upsilon (\hat x)$ has simple zeros on $\gR$ at each of the 
points $\eta_{j,l}$, although it may seem surprising. Moreover, 
the function
\[
\frac{1}{\Upsilon^2(\hat x)} = \frac{1}{(x-\eta_{1,1})(x-\eta_{2,1})
(x-\eta_{1,2})(x-\eta_{2,2})}
\] 
has double poles at each $\eta_{j,l}$ as a function on $\gR$, 
while the same function, but considered on $\overline{\mathbb{C}}$, 
has simple poles at those points.
Function $\Upsilon(\hat x)$ has double poles at the infinities on~$\gR$ and on~$\gR_3$. 

A meromorphic function on $\gR$ or on $\gR_3$ (or on any other compact Riemann surface with a structure of complex manifold) 
is a function single-valued on 
the Riemann surface and 
having a finite number of singularities, each of which is 
a pole of finite order.
  
A differential 1-form on a complex manifold is 
as a set of expressions $f_s (\tau_s)\, d\tau_s$ (each defined in $\mathcal{U}_s$), such that 
\begin{equation}
\frac{f_j}{f_s} = \frac{d\tau_s}{d\tau_j} 
\label{eq211y}
\end{equation}
in $\mathcal{U}_j \cap \mathcal{U}_s$.
The form is analytic in some domain (or everywhere) if $f_j$ is analytic there (or if all $f_j$ are analytic).
It is important for our consideration that  
the form $dx / \Upsilon (\tilde x)$ is analytic {\em everywhere\/} in~$\gR_3$.  

Poles and zeros of forms and functions are defined also in the local variables. 
A residue of a 1-form is defined invariantly, i.~e.\ the residue does not depend on the 
choice of the local variable. Namely, a 1-form has residue equal to $a$ at some point $\tau_j'$ if 
it can be locally represented using $\tau_j$ as 
\[
\left( \frac{a}{\tau_j - \tau_j'} + c(\tau_j) \right) d\tau_j, 
\] 
where $c$ is some function regular near $\tau_j'$.
Note that the residue of a function is not invariant.

Below we use  
contour integration on the Riemann surface 
$\gR_3$ based on the structure of a complex manifold on $\gR_3$. 
This integration  
is introduced also locally. Let a 1-form $f_j d\tau_j$ 
and an oriented contour $\gamma$ on $\gR_3$ be given. 
In each neighborhood $\mathcal{U}_j$ 
passed by $\gamma$ one can define a term 
\[
\int_{\gamma_j} f_j \, d\tau_j,
\]
where $\gamma_j$ is some part of $\gamma \cap \mathcal{U}_j$. 
The total integral is a sum of all such parts provided that 
the concatenation of all $\gamma_j$ is~$\gamma$. 
The condition (\ref{eq211y}) guarantees that integration 
is invariant with respect to the choice of the local variables. 
The Cauchy's theorem is inherited from usual complex analysis in a trivial way 
(in each $\mathcal{U}_j$). 
The theorem states that the contour of integration can be freely deformed in some domain on $\gR_3$, 
provided the integrand (a 1-form) is analytic in this domain. The result of integration remains unchanged 
under this deformation. 

Introduce a Riemann surface over a complex manifold. Namely, let $\mathcal{R}$ be a complex manifold described above, 
$\mathcal{R}'$ be a manifold of real dimension~2, and 
$P: \mathcal{R}' \to \mathcal{R}$ be a continuous mapping. $\mathcal{R}'$ is a Riemann surface
if all $P^{-1} (\mathcal{U}_s)$, $\mathcal{U}_s \subset \mathcal{R}$ have structure of a Riemann surface.
Using this definition, one can show that $\gR_3$ is a 3-sheet Riemann surface over~$\gR$.
Besides, one can see that $\gR_3$ over~$\gR$ is a Riemann surface without branch points. 
To prove this, one can consider the neighborhoods of $\eta_{j,l}$ in corresponding 
local variables.

  
\subsection{The main result of \cite{Shanin2020} (the Sommerfeld integral)}
\label{wedge2b}

Let $\tilde x_1, \dots \tilde x_4$ be the points of $\gR_3$ defined by
\[
\tilde x_1 = \tilde X(x_{\rm in} , 3),
\quad 
\tilde x_2 = \tilde X(x_{\rm in} , 4),
\quad 
\tilde x_3 = \tilde X(x_{\rm in}^{-1} , 6),
\quad 
\tilde x_4 = \tilde X(x_{\rm in}^{-1}, 1).
\]
One can see that 
\[
y(\tilde x_1) = y (\tilde x_4) = y_{\rm in}, 
\quad 
y(\tilde x_2) = y (\tilde x_3) = y^{-1}_{\rm in}.
\]
It is shown in \cite{Shanin2020} that 
the point $\tilde x_1$ corresponds to the incident wave, and three other points
correspond to the reflected waves of $\gS_3$ (this can be seen from the ``wavenumbers'' $x$ and $y$). Note that the choice of sheets for these points is made according to contours of integration in the Sommerfeld integral (see integration contours in Fig.~\ref{fig03} and Theorem~\ref{th_previous} below) 

Following \cite{Shanin2020}, let us formulate the functional problem for the function $A(\tilde x)$, 
which is the Sommerfeld transformant of the total wave (see below): 

\vskip 6pt
\noindent
{\bf Functional problem~1. }
{\em

a) The function $A(\tilde x)$ should be meromorphic on~$\gR_3$. 

b) Function $A(\tilde x)$
should have four poles on~$\gR_3$
at the points $\tilde x_1$, $\tilde x_2$, $\tilde x_3$, $\tilde x_4$ of 
order~1, and no other poles.

c) 
The residues of $A(\tilde x)$ at the poles are specified by setting the residues of the 
poles of the form $A(\tilde x) dx / \Upsilon (\hat x)$. 
Namely, this form should have the residues $-(2\pi i)^{-1}$ at the 
points $\tilde x_1$ and $\tilde x_3$, and the 
residues $(2\pi i)^{-1}$ at the points $\tilde x_2$ and~$\tilde x_4$. 
}

Some comments should be made. 
According to the first condition, 
$A(x)$ should be an analytic function, 6-valued on $\overline{\mathbb{C}}$. 
It should have  branch points of order two 
at $\eta_{1,1}$, $\eta_{1,2}$, $\eta_{2,1}$, $\eta_{2,2}$.
The sheets of the Riemann surface of $A$ should be linked in the way shown in Fig.~\ref{fig03}.
Function $A(\tilde x)$ should be regular as $|x|\to \infty$ 
on each of the six sheets of~$\gR_3$. 

The residues of the function $A(\tilde x)$ can be specified, 
provided $x$ is taken as a local variable near the poles. The residues 
at $\tilde x_j$, $j = 1, \dots , 4$ are equal to 
\[
-(2\pi i)^{-1} Y_1 , \qquad (2\pi i)^{-1} Y_2 , \qquad -(2\pi i)^{-1} Y_3, \qquad (2\pi i)^{-1} Y_4,   
\]  
where 
\[
Y_j = \Upsilon (\hat x_j).
\]
One can see that $Y_2 = -Y_1$, $Y_4 = - Y_3$. 

Obviously, Functional problem~1 defines $A$ up to an arbitrary additive constant (at least). 
Moreover: 

\begin{proposition}
\label{prop_uniqueness}
If $A(\tilde x)$ and $A'(\tilde x)$ are different solutions of Functional problem~1, 
then $A(\tilde x) - A' (\tilde x)$ is a constant.
\end{proposition} 
To prove this proposition, note that the function $A(\tilde x) - A' (\tilde x)$ is regular everywhere on 
$\gR_3$, thus it is constant (by Liouville's theorem for Riemann surfaces, see \cite{Koch1991b} chapter 13). 



Function $A$, formally, 
has been found in \cite{Shanin2020} (see (70) there), 
and it has been expressed in elliptic functions, and this solution is hardly 
practical. However, the formulation of the problem is of algebraic nature, 
so one can expect a purely algebraic solution.  

The main result of \cite{Shanin2020} related to the problem of diffraction
in the domain shown in Fig.~\ref{fig12b}
can be formulated as a theorem: 

\begin{theorem}
\label{th_previous}

Let  the function $A(\tilde x)$ obey Functional problem~1. 
Introduce the contours 
\begin{equation}
\Gamma_2 = J_3 + J_2 + J_1' + J'_4,  
\label{eq213}
\end{equation} 
\begin{equation}
\Gamma_3 =  J_2 + J_1' + J'_4 + J'_3, 
\label{eq214}
\end{equation} 
where the
contours $J_3$, $J_2$, $J_1'$, $J'_4$, $J'_3$ on $\gR_3$ are shown in Fig.~\ref{fig03}. 

Define the functions $u_j(m,n)$, $m,n \in \mathbb{Z}$, $j = 2,3$, by the Sommerfeld integral: 
\begin{equation}
u_j (m,n) = \int_{\Gamma_j}
w_{m,n} (\hat x , y(\hat x)) A(\tilde x) \frac{dx}{\Upsilon(\hat x)}. 
\label{eq212}
\end{equation}  

Then 

a) $u_2 (m,n) = u_3 (m,n)$ for $m,n \le 0$. 


b) The function 
\[
u(m,n) = \left\{ \begin{array}{ll}
u_2(m,n) , & m \le 0 , \quad n > 0 \\
u_3(m,n) , & n \le 0 , \quad m > 0 \\
u_2(m,n) = u_3(m,n), & m \le 0 , \quad n \le 0 
\end{array} \right.
\] 
defined  in the domain shown in Fig.~\ref{fig12b}
obeys  the diffraction problem formulated 
in Subsection~\ref{wedge1}.   

\end{theorem}

The proof can be found in \cite{Shanin2020}. An alternative proof is given in Appendix~1. 

\vskip 6pt
\noindent
{\bf Remark.} 
The notation for the contours $\Gamma_2$ and $\Gamma_3$ is kept similar to that of~\cite{Shanin2020}, 
where a family of Sommerfeld contours on a torus is introduced.
Totally, one can define 12 contours $\Gamma_1 \dots \Gamma_{12}$, 
describing  $u(m,n)$ on the whole branched surface $\gS_3$. 
Contours $J_2$ and $J'_1$ encircle the point $x = 0$ on corresponding 
sheets. Contours $J_3$, $J'_4$, $J'_3$ encircle the inifinities on corresponding sheets.

According to the formula (\ref{eq212}), adding an arbitrary constant to $A$ does not 
change the integral. 

The main result of this section is as follows:  to find the solution of the 
diffraction problem, one should first solve the Functional problem~1, and then 
substitute $A(\tilde x)$ into the Sommerfeld integral~(\ref{eq212}).


\section{Mathematical basics of solving the functional problem for $A(x)$}
\label{sec_math_bas}

\subsection{Symmetries of the Riemann surface }
\label{subsec_sym}

Consider the cyclical substitution of sheets  
\begin{equation}
1 \to 3 \to 5 \to 1 ,
\qquad 
2 \to 4 \to 6 \to 2 . 
\label{eq_cyc}
\end{equation}
This substitution of sheets generates a symmetry (referred to as $\Lambda$) of $\gR_3$, Namely,
\begin{equation}
\Lambda ( \tilde X(x , j )) = \tilde X (x ,j'),
\label{eq:La}
\end{equation} 
where  $j'$ is obtained from $j$ by applying the mapping (\ref{eq_cyc}). 

One can see that if a function $f(\tilde x)$ is meromorphic on $\gR_3$ then 
the same is valid for $f(\Lambda(\tilde x))$. Moreover, 
since $P_{3:1}(\Lambda (\tilde x)) = P_{3:1}(\tilde x)$,
$\Lambda$ does not change the value 
of $\Upsilon(\tilde x)$:
\begin{equation}
\Upsilon (\Lambda (\tilde x)) =  \Upsilon (\tilde x). 
\label{eq306b}
\end{equation}  

Another symmetry (referred to as $\Pi$) is defined on $\gR$ and on $\gR_3$ as follows: 
\begin{equation}
\Pi(\hat X(x , j)) = \hat X ( x , 3-j), 
\qquad
\Pi(\tilde X(x , j)) = \tilde X ( x , 7-j). 
\label{eq305}
\end{equation}
One can see that  
if a function $f(\tilde x)$ is meromorphic on $\gR_3$ then 
the same is valid for $f(\Pi(\tilde x))$.
A direct check shows that 
\begin{equation}
\Upsilon (\Pi (\hat x)) = - \Upsilon (\hat x).
\qquad 
\Upsilon (\Pi (\tilde x)) = - \Upsilon (\tilde x).
\label{eq306a}
\end{equation}

Obviously, 
\[
\Pi (\Pi (\hat x)) = \hat x, 
\qquad 
\Pi (\Pi (\tilde x)) = \tilde x, 
\qquad
\Lambda( \Lambda( \Lambda(\tilde x))) = \tilde x. 
\]

The mappings $\Lambda$ and $\Pi$ are {\em desk transformations\/} of $\gR_3$ in terms 
of~\cite{Khovanskii2014}. 

The symmetries $\Pi$ and $\Lambda$ can be used to simplify the formulation 
of the functional problem for $A(\tilde x)$. These simplifications can be formulated 
as the following propositions. 

\begin{proposition}
\label{prop_sym_Pi}
If $A$ is a solution of the functional problem formulated above, then 
\begin{equation}
A(\Pi (\tilde x)) = A(\tilde x).
\label{eq306}
\end{equation} 
\end{proposition}

To prove this proposition, note that the poles of $A$ have the property 
$\Pi (\tilde x_1) = \tilde x_2$, $\Pi (\tilde x_3) = \tilde x_4$, 
moreover, the residues at corresponding poles are the same. 
Thus, $A(\Pi(\tilde x))$ also obeys the functional problem. 
Taking into account Proposition~\ref{prop_uniqueness}, and the fact that at some branch points 
$\Pi (\tilde x) = \tilde x$, obtain the result.  

\begin{proposition}
\label{prop_Lambda}
Let $A$ be a solution of Functional problem~1. Then 
it can be represented as a sum of 3 components:  
\begin{equation}
A(\tilde x) = A_0 (\tilde x) + A_1 (\tilde x) + A_2(\tilde x),
\label{eq306m}
\end{equation}
where $A_1$, $A_2$, $A_3$ are meromorphic on $\gR_3$, and  
\begin{equation}
A_0 (\Lambda (\tilde x)) = A_0 (\tilde x),
\quad 
A_1 (\Lambda (\tilde x)) = \varpi A_1 (\tilde x),
\quad  
A_2 (\Lambda (\tilde x)) = \varpi^{-1} A_2(\tilde x)
\label{eq306n}
\end{equation}
for all $\tilde x \in \gR_3$, 
\begin{equation}
\varpi \equiv e^{2\pi i / 3}. 
\label{eq014x}
\end{equation}
\end{proposition}

The proof is given by explicit formulae: 
\begin{equation}
A_0 (\tilde x) = \frac{1}{3} \left( 
A(\tilde x) + A(\Lambda (\tilde x)) + A(\Lambda (\Lambda(\tilde x)))
\right) ,
\label{FT1}
\end{equation}
\begin{equation}
A_1 (\tilde x) = \frac{1}{3} \left( 
A(\tilde x) + \varpi^2 A(\Lambda (\tilde x)) + \varpi A(\Lambda (\Lambda(\tilde x)))
\right) ,
\label{FT2}
\end{equation}
\begin{equation}
A_2 (\tilde x) = \frac{1}{3} \left( 
A(\tilde x) + \varpi A(\Lambda (\tilde x)) + \varpi^2 A(\Lambda (\Lambda(\tilde x)))
\right) .
\label{FT3}
\end{equation}
Indeed, (\ref{FT1}), (\ref{FT2}), (\ref{FT3}) constitute a discrete Fourier transform on 
each set $\{ \tilde x , \Lambda(\tilde x) , \Lambda(\Lambda(\tilde x)) \}$.

By construction, 
the functions $A_0(\tilde x)$, $A_1(\tilde x)$, $A_2(\tilde x)$ obey the following functional problem: 

\vskip 6pt
\noindent  
{\bf Functional problem~2.}

{\em 
a)  
Functions $A_0$, $A_1$, $A_2$ should be meromorphic on $\gR_3$. 

b)
Each of the functions 
$A_0$, $A_1$, $A_2$
is allowed to have 12 poles located at $\tilde x_j$, $\Lambda (\tilde x_j)$, $\Lambda(\Lambda (\tilde x_j))$,
$j \in \{ 1,2,3,4\}$.  All poles should be of order~1.

c)
All residues of the poles of the functions $A_0$, $A_1$, $A_2$ are
prescribed. For each of these functions
they are equal 
to $-(6\pi i)^{-1} Y_1$ at the points $\tilde x_1, \tilde x_2$, and to   
$-(6\pi i)^{-1} Y_3$ at the points $\tilde x_3, \tilde x_4$. 
The residues at the poles $\Lambda (\tilde x_j)$ and $\Lambda (\Lambda (\tilde x_j))$
can be found from these values and the relations (\ref{eq306n}). 

d) 
Functions $A_0$, $A_1$, $A_2$
should obey relations (\ref{eq306n}).
}

Similarly to Proposition~\ref{prop_uniqueness}, the solution of this functional problem is unique 
for $A_1$, $A_2$, and defined up to an additive constant for~$A_0$.


\subsection{Fields of functions meromorphic on Riemann surfaces}  
\label{ss:fields}

Above, we considered {\em wave fields\/}, i.~e.\ solutions of a (discrete) Helmholtz equation. 
Here and below we consider {\em algebraic fields\/} that are sets of elements, on which 
the arithmetic operations are defined and possess usual properties.
We hope that the usage of the term ``field'' should not cause a confusion. 
In more details, we are going to study functional fields, i.~e.\ the elements of fields  are some functions. 
A set of functions is a field if a sum, a product, a difference, or a ratio of two elements
of the set belongs to this set. Indeed, in the last case the denominator should not be identically equal 
to zero.    

It is obvious that the following proposition is correct ( see \cite{Koch1991b}, chapter 11): 

\begin{proposition}
\label{prop_field_1}
Let $\mathcal{R}$ be a compact Riemann surface with a structure of complex manifold defined on it. 
Then the set of all functions  meromorphic on $\mathcal{R}$ is a field.  
\end{proposition}

We will use three Riemann surfaces as $\mathcal{R}$, namely $\overline{\mathbb{C}}$,
$\gR$, and $\gR_3$. 
The fields of functions meromorphic on them  will be denoted by $\gK_0$, $\gK_1$, $\gK_3$. We are particularly interested in 
the field $\gK_3$, since $A(\tilde x) \in \gK_3$.

Let $\mathcal{R}$ be a Riemann surface over the Riemann surface $\mathcal{R}'$ (both compact), 
and let $P$ the corresponding projection 
$P: \quad \mathcal{R} \to \mathcal{R}'$. Let the fields of meromorphic 
functions on $\mathcal{R}$ and $\mathcal{R}'$ be denoted by $\gK$ and~$\gK'$, respectively. 
We say that $f \in \gK$ belongs also to $\gK'$ if $f$ is single-valued on $\gK'$, 
i.~e.\ if there exists a function $f' \in \gK'$ such that 
$f(x) = f'(P(z))$, $x \in \mathcal{R}$. 
In the same sense, any function $f' \in \gK'$ belongs to~$\gK$, since 
$f(x) = f'(P(z))$ is a definition of an appropriate function $f (x)$. The existence of 
continuous mappings $P_{3:1}$ and $P_{1:0}$ leads to the 
following statement: 

\begin{proposition}
\label{prop_field_2}
The following inclusions are valid: 
\[
\gK_0 \subset \gK_1 \subset \gK_3.
\]
In the usual terms (see, for example, \cite{EmilArtin1997}), $\gK_1$ is an extension of $\gK_0$, and $\gK_3$ is extension of $\gK_1$.
\end{proposition}

\vskip 6pt
\noindent 
{\bf Remark.}
Mappings 
\begin{equation}
f(\hat x) \to f(\Pi (\hat x)), \qquad f(\hat x) \in \gK_1,
\label{eq_auth1}
\end{equation} 
\begin{equation}
f(\tilde  x) \to f(\Lambda (\tilde x)) , \qquad f(\tilde x) \in \gK_3
\label{eq_auth2}
\end{equation}
are authomorphisms of $\gK_1$ and $\gK_3$, respectively.  
As it is common for the Galois theory \cite{Khovanskii2014}, 
the following proposition links field authomorphisms and 
extensions:

\begin{proposition}
\label{pr_auth} 
The elements of $\gK_1$ invariant with respect to (\ref{eq_auth1}) belong to~$\gK_0$, 
The elements of $\gK_3$ invariant with respect to (\ref{eq_auth2}) belong to~$\gK_1$.
\end{proposition}

Indeed, these facts are obvious for the surfaces under consideration.  

The field $\gK_0$ 
of functions $f(x)$ meromorphic on the Riemann sphere $\mathbb{\bar C}$  
consists of all rational functions of~$x$ (a rational function is a ratio of two polynomials). 
For the field $\gK_1$, the following proposition is valid: 

\begin{proposition}
\label{prop_K1}
Each element of $z \in \gK_1$ can be uniquely represented as
\begin{equation}
z(\hat x) = q_0 (x) + q_1(x) \Upsilon(\hat x),
\qquad 
q_0, q_1 \in \gK_0.
\label{eq103x}
\end{equation}
\end{proposition}

\noindent
{\bf Proof.}
Let be $x \in \overline{\mathbb{C}}$. 
The functions 
\[
\frac{z(\hat x) + z(\Pi(\hat x))}{2} 
\quad \mbox{and} \quad 
\frac{z(\hat x) - z(\Pi (\hat x))}{2\Upsilon (\hat x )}
\]
are meromorphic on $\overline{\mathbb{C}}$ since they are invariant with 
respect to (\ref{eq_auth1}).  
They are $q_0$ and $q_1$, respectively.
The elements with $q_1 \equiv 0$ are the elements of $\gK_1$
that belong to~$\gK_0$.  $\square$  

\vskip 6pt

One can see that $\gK_1$ is an {\em algebraic extension\/} of $\gK_0$, i.~e.\ an irrationality 
$\Upsilon (x)$ is added to the field $\gK_0$. 
The irrationality is a solution of the algebraic equation of order~2, 
whose coefficients belong to $\gK_0$:
\begin{equation}
\Upsilon^2  
- f= 0 , 
\qquad f(x) = (x - \eta_{1,1})(x - \eta_{1,2})(x - \eta_{2,1})(x - \eta_{2,2}).
\label{eq101} 
\end{equation}
One can see that (\ref{eq103x}) is an expansion of the form 
\begin{equation}
z(\hat x) = q_0(x)\, \omega_0(\hat x) + \dots + q_{j-1}(x)\, \omega_{j-1}(\hat x),
\label{eq102}
\end{equation}
where $j = 2$; $q_0, q_1 \in \gK_0$, and the set  
\begin{equation}
\Omega_{1:0} \equiv [\omega_0 , \dots, \omega_{j-1}] = [1 , \Upsilon(\hat x)]
\label{eq103}
\end{equation}
is the {\em basis of the extension}. 
In \cite{Koch1991b}, chapter 23
it is shown that such a basis always exits.
The number $j$ of elements of the basis is referred to as the {\em degree\/} of extension of 
$\gK_1$ over~$\gK_0$. This degree is equal to~2, and it is the same as number of sheets 
of $\gR$ over $\overline{\mathbb{C}}$. 

Consider the field $\gK_3$, i.~e.\ the field of functions meromorphic on $\gR_3$. 
The following statement is valid: 

\begin{theorem}
\label{th_K3}
Let there 
exist non-zero functions $F_1(\tilde x) , F_2 (\tilde x) \in \gK_3$
having  properties 
\begin{equation}
F_1 (\Lambda(\tilde x)) = \varpi F_1 (\tilde x),  
\label{eq_prop1}
\end{equation} 
\begin{equation}
F_2 (\Lambda(\tilde x)) = \varpi^{-1} F_2 (\tilde x).   
\label{eq_prop2}
\end{equation}
Then for any function 
$z(\tilde x) \in \gK_3$ there exists a (unique) representation
\begin{equation}
f(\tilde x) = q_0 (\hat x) + q_1 (\hat x) F_1 (\tilde x) + q_2 (\hat x) F_2 (\tilde x),
\label{eq106x}
\end{equation}
where $q_0 , q_1, q_2 \in \gK_1$.

The functions $F_1(\tilde x)$ and $F_2(\tilde x)$ 
are solutions of cubic equations
with coefficients belonging to $\gK_1$:
\begin{equation}
F_1^3 - G_1  =0 , 
\qquad 
F_2^3 - G_2  =0 , 
\qquad 
G_{1,2} \in \gK_1.
\label{eq106y}
\end{equation}
\end{theorem} 

Indeed, this statement means that the extension $\gK_3$ over $\gK_1$ has a basis 
\begin{equation}
\Omega_{3:1} = [1, F_1 , F_2]
\label{eq106}
\end{equation}
The functions $F_{1,2}$ should be single-valued on $\gR_3$, three-valued on $\gK_1$, and six-valued on 
$\overline{\mathbb{C}}$. 

\noindent 
{\bf Proof. } a) 
Consider the combinations similar to (\ref{FT1}), (\ref{FT2}), (\ref{FT3}):  
\begin{equation}
f_0 (\tilde x) = \frac{1}{3}
\left( f (\hat x) + f (\Lambda(\hat x)) + f (\Lambda(\Lambda(\hat x))) \right),  
\label{eqa01}
\end{equation}
\begin{equation}
f_1 (\tilde x) = \frac{1}{3}
\left( f (\hat x) + \varpi^{-1} f (\Lambda(\hat x)) + 
\varpi f (\Lambda(\Lambda(\hat x))) \right),  
\label{eqa02}
\end{equation}
\begin{equation}
f_2 (\tilde x) = \frac{1}{3}
\left( f (\hat x) + \varpi f (\Lambda(\hat x)) 
+ \varpi^{-1} f (\Lambda(\Lambda(\hat x))) \right),  
\label{eqa03}
\end{equation}
Obviously, if $f_0$, $f_1$, $f_2$ are known, the function $f$
can be reconstructed by 
\begin{equation}
f (\tilde  x) =  f_0 (\tilde  x) + f_1 (\tilde x) + f_2 (\tilde x) . 
\label{eqa04}
\end{equation}

Note that 
\begin{equation}
f_0 (\Lambda(\tilde x)) = f_0 (\tilde x), 
\qquad 
f_1 (\Lambda(\tilde x)) = \varpi f_1 (\tilde x), 
\qquad 
f_2 (\Lambda(\tilde x)) = \varpi^{-1} f_2 (\tilde x). 
\label{eqa08}
\end{equation}
According to Proposition~\ref{pr_auth}, $f_0 \in \gK_1$, and one can take $q_0 = f_0$. 
The coefficients $q_1$ and $q_2$ are chosen as 
\[
q_1 = f_1 (\tilde x) / F_1 (\tilde x), 
\qquad 
q_2 = f_2 (\tilde x) / F_2 (\tilde x). 
\]
These functions are invariant with respect to (\ref{eq_auth1}), 
thus they belong to $\gK_1$. The uniqueness of the representation (\ref{eq106x}) 
(provided the functions $F_1$ and $F_2$ are fixed)
follows from the construction of the coefficients. 

b) Due to (\ref{eq_prop1}) and (\ref{eq_prop2}), the functions $G_1 = (F_1)^3$ and $G_2 = (F_2)^3$ 
are invariant with respect to (\ref{eq_auth2}), thus, 
$F_{1,2} \in \gK_1$. $\square$

\vskip 6pt

The functions $F_1$ and $F_2$ do exist, they will be built explicitly below. A corollary of 
Theorem~\ref{th_K3} is that the Sommerfeld transformant $A$ has a representation 
\begin{equation}
A(\tilde x) = q_0(\hat x) + q_1(\hat x) F_1(\tilde  x) + q_2(\hat x) F_2(\tilde x) , 
\label{eq107}
\end{equation}
where $q_j(\hat x)$ belong to $\gK_1$ being rational functions of 
$x$ and $\Upsilon(\hat x)$. 
Comparing formulae (\ref{FT1})--(\ref{FT3}) with (\ref{eqa01})--(\ref{eqa03})
we conclude that 
\begin{equation}
A_0(\tilde x) = q_0(\hat x),
\qquad 
A_1(\tilde x) = q_1(\hat x) F_1(\tilde x),
\qquad 
A_2(\tilde x) = q_2(\hat x) F_2(\tilde x).
\label{eq107a}
\end{equation}

Finding the functions $F_1(\tilde x)$ and $F_2(\tilde x)$ is an unusual problem since
no function whose Riemann surface is $\gR_3$ is given {\em a~priori}.
Finding the coefficients  $q_j (x)$ is, conversely, an almost trivial  
task when the basis (\ref{eq106}) is built. 
They are constructed by using the knowledge of poles and residues 
of the Sommerfeld transformant~$A$.  

The choice of functions $F_1$ and $F_2$ is not unique. Each of them can be multiplied by any nonzero element of 
$\gK_1$, still keeping the properties (\ref{eq_prop1}), (\ref{eq_prop2}). Below we are trying to construct the 
functions having the simplest structure, i.~e.\ having as small amount of poles / zeros on~$\gR_3$ as possible. 

\vskip 6pt
\noindent
{\bf Remark.} Consider the expansion (\ref{eq107}). The basis functions $F_1$ and $F_2$
are constructed below depending on~$K$ and not depending on $x_{\rm in}$.
The coefficients $q_0$, $q_1$, $q_2$, conversely,  do depend on $x_{\rm in}$. 
The structure of functions $F_{1,2}$ guarantee the validity of conditions~a) and~d) of the Functional problem~2,
while the choice of the coefficients $q_{0,1,2}$ guarantee the conditions b) and~c).



\subsection{Abelian integral of the first kind on $\gR$}

Introduce the Abelian integral of the first kind on~$\gR$.
A detailed description of this subject can be found, e.g., in
\cite{Koch1991b}, chapter~12. 
Since $\gR$ is a torus, a surface of genus~1, 
there is one Abelian integral analytic everywhere 
(indeed, defined up to a constant factor and a constant additive term). 
This Abelian integral is an integral of a differential 1-form, 
which is analytic everywhere on~$\gR$. As we have mentioned, 
$dx / \Upsilon$ is such a form, thus the Abelian integral of the 
first kind is 
\begin{equation}
\chi(\hat x) = \int \limits_{\eta_{2,1}}^{\hat x} \frac{dx'}{\Upsilon(\hat x')}. 
\label{eq005}
\end{equation}
The choice of the starting point  $\eta_{2,1}$  is arbitrary.
The integral is assumed to be taken along some oriented contour $\gamma$ on $\gR$
connecting $\eta_{2,1}$ and $\hat x$: $\chi (\hat x) = \chi(\hat x , \gamma)$.
Note that (\ref{eq005}) is an Abelian integral on $\gR_3$ as well.
 

The integrals (\ref{eq005}) taken along the closed contours
$\sigma_\alpha$ and $\sigma_\beta$ on $\gR$ are 
{\em the periods\/} of $\chi(\hat x)$ referred to as 
$T_{\alpha}$ and $T_{\beta}$:
\begin{equation}
T_{\alpha} = \int_{\sigma_{\alpha}} \frac{dx}{\Upsilon (\hat x)} ,
\qquad 
T_{\beta} = \int_{\sigma_{\beta}} \frac{dx}{\Upsilon (\hat x)}.
\label{eq005b}
\end{equation}

The function $\chi$ is used below as a multiple-valued mapping between $\gR$ and the complex plane of~$\chi$.
%
%
Introduce also an inverse mapping $\psi: \, \chi \to \hat x$. 
We will use the following properties of the mappings $\chi$ and $\psi$ 
that can be found in any textbook on elliptic functions, e.g.\ \cite{Koch1991b}. 
 
\begin{proposition}
\label{prop_el}
a) Let $\gamma_0$ be some contour connecting $\eta_{2,1}$ with $\hat x$. Then all values of $\chi(\hat x , \gamma)$ 
are $ \chi (\hat x , \gamma_0) + j T_{\alpha} + l T_{\beta}$, $j,l \in \mathbb{Z}$.

b) Mapping $\psi$ is a bijection of $\gR$ and the elementary
parallelogram 
with vertexes $(0, T_\alpha , T_\beta + T_\alpha,  T_\beta)$
and with the opposite sides glued together. 
\end{proposition}  
 
As it follows from this proposition, mapping $\psi$ is defined correctly and 
is bi-periodic:  
\begin{equation}
\psi (\chi + T_{\alpha}) = \psi (\chi + T_\beta) = \psi (\chi).
\label{eq005d}
\end{equation}

\begin{figure}[ht]
\centering
\includegraphics[width=0.8\textwidth]{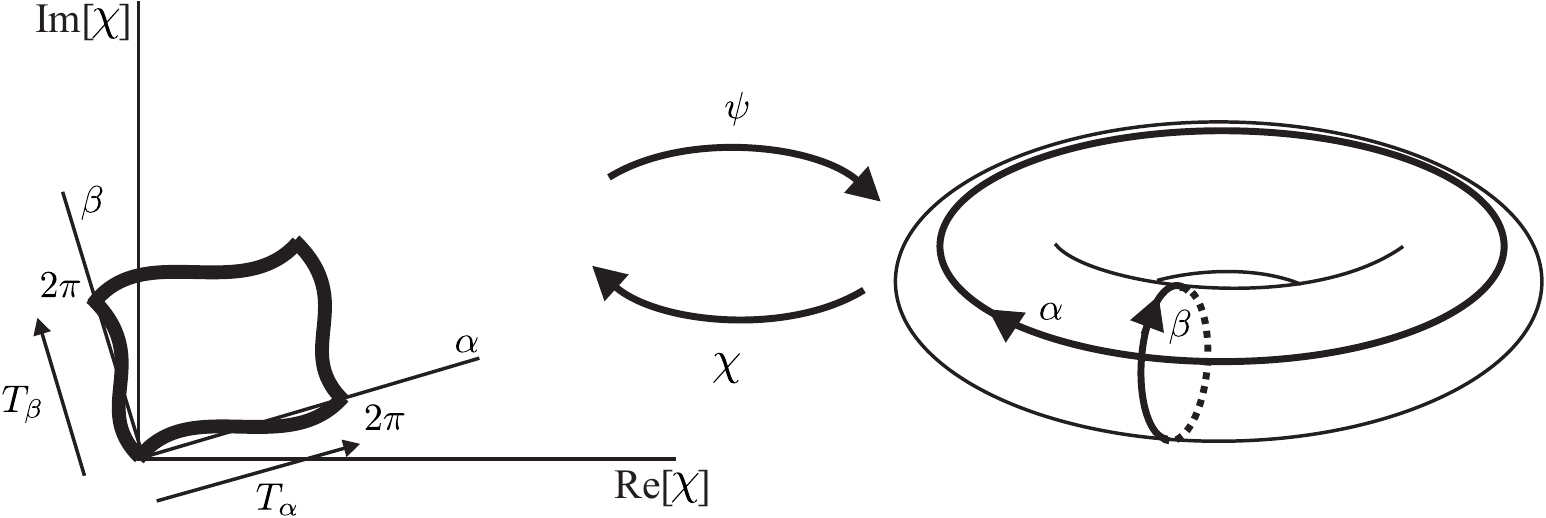}
\caption{Elementary parallelogram in the $\chi$-plane and coordinates $(\alpha,\beta)$
on a torus} 
\label{fig23}
\end{figure}

Being cut along the contours $\sigma_\alpha$ and $\sigma_\beta$,
the surface $\gR$ becomes (topologically) a parallelogram, as it is shown in 
Fig.~\ref{fig23}.

Proposition~\ref{prop_el} can be used for introduction of coordinates 
$\alpha$ and $\beta$ on $\gR$, revealing the structure of $\gR$ as the structure of a torus. Namely, the coordinates $\alpha$ and $\beta$ can be introduced as
linear combinations 
\begin{equation}
\alpha = c_{1,1} {\rm Re}[\chi] + c_{1,2} {\rm Im}[\chi],
\qquad  
\beta  = c_{2,1} {\rm Re}[\chi] + c_{2,2} {\rm Im}[\chi],
\label{eq005e}
\end{equation}
with the coefficients $c_{j,k}$ found from the following equations
\begin{equation}
c_{1,1} {\rm Re}[T_\alpha] + c_{1,2} {\rm Im}[T_\alpha] = 2\pi,
\qquad  
c_{1,1} {\rm Re}[T_\beta ] + c_{1,2} {\rm Im}[T_\beta]  = 0,
\label{eq005f}
\end{equation}
\begin{equation}
c_{2,1} {\rm Re}[T_\alpha] + c_{2,2} {\rm Im}[T_\alpha] = 0,
\qquad  
c_{2,1} {\rm Re}[T_\beta ] + c_{2,2} {\rm Im}[T_\beta]  = 2\pi.
\label{eq005g}
\end{equation}

The coordinates $(\alpha, \beta)$ on $\gR$ are shown in 
Fig.~\ref{fig23}, right. The surface $\gR$ is displayed schematically as a torus, i.~e.\
$\gR$ is deformed in an appropriate way. The resulting surface is compact, 
thus, the infinities are represented as two points on it.  
The coordinate lines of $\alpha$ and $\beta$ on the initial representation of $\gR$
are close\footnote{
Coordinates $\alpha$ and $\beta$ are close to the coordinates 
$\alpha$ and $\beta$ defined in~\cite{Shanin2020}, but not exactly the same. Note that the requirement that $\beta = \pi$
on the ``real waves'' line is not fulfilled in the new formulation.}
to those shown in Fig.~4 of 
\cite{Shanin2020}.


The torus $\gR$ corresponds to the parallelogram 
\[
\gR : \quad  0 \le \alpha < 2\pi, \quad  0 \le \beta < 2\pi  ,
\]
while, 
according to Proposition~\ref{prop_cont},
the torus $\gR_3$ corresponds to the parallelogram 
\[
\gR_3 : \quad  0 \le \alpha < 6\pi, \quad  0 \le \beta < 2\pi  .
\]
Each point $(\alpha, \beta) \in \gR$  has three preimages $P_{3:1}^{-1} (\alpha, \beta)$:
$
(\alpha, \beta),(\alpha+ 2\pi, \beta),(\alpha+ 4\pi, \beta)  
$
on~$\gR_3$.
 
The symmetries $\Lambda$ and $\Pi$ have the following representations in the 
coordinates $(\alpha, \beta)$: 
\[
\Lambda: \quad \alpha \to \alpha + 2\pi , \quad \beta \to \beta ,
\]
\[
\Pi: \quad \alpha \to 4\pi - \alpha , \quad \beta \to -\beta. 
\]


\section{Finding the basis functions $F_1$, $F_2$ }
\label{sec_basis}


\subsection{Elementary meromorphic functions on $\gR$}

Our aim is to build functions $F_{1,2}$. For this, we will use an auxiliary function
\[
M(\hat x) = M(\hat b_1 , \hat b_2 , \hat a_1 , \hat a_2 ; \hat x) 
\]
($\hat x \in \gR$ 
is a variable, $\hat b_{1,2}, \hat a_{1,2} \in \gR$ are parameters) such that:  
$M \in \gK_1$, it has poles only at $\hat b_{1,2}$, the poles are simple  
if $\hat b_1 \ne \hat b_2$ or double if 
$\hat b_1 = \hat b_2$, and it has zeros only at $\hat a_{1,2}$ (simple zeros if $\hat a_1 \ne \hat a_2$
or a double zero if $\hat a_1 = \hat a_2$).

Indeed, each such function (if it exists) is defined up to a constant factor. This factor is not important for us 
and we suppress it in the Ansatzes written below.  

Such a function exists not for any set $(\hat a_1 , \hat a_2 , \hat b_1 , \hat b_2)$.
A criterion of existence of such a function $M$ is known:

\begin{theorem}
\label{th_Abel}
Function $M(\hat b_1 , \hat b_2 , \hat a_1 , \hat a_2 ; \hat x)$ exists
iff there exist oriented contours $\gamma_1$ and $\gamma_2$
on $\gR$, 
such that 
$\gamma_1$ goes from $\hat a_1$ to $\hat b_1$, 
$\gamma_2$ goes from $\hat a_2$ to $\hat b_2$, and 
\begin{equation}
\int \limits_{\gamma_1}\frac{dx}{\Upsilon(\hat x)} + 
\int \limits_{\gamma_2}\frac{dx}{\Upsilon(\hat x)}
 = 0 .
\label{eq006}
\end{equation}
\end{theorem}

This is a particular case of 
the Abel's theorem (see \cite{Springer2001}, chapter~10).
The criterion has a transcendent (non-algebraic) character.
Surprisingly, it is possible to formulate an algebraic criterion of existence 
of such a function. Below we formulate a set of propositions describing different 
choices of $\hat a_{1,2} , \hat b_{1,2}$. 
We formulate two cases important for our consideration 
as Propositions~\ref{prop_M3} and~\ref{prop_M4} here, 
and, for completeness, formulate two more cases  as 
Propositions~\ref{prop_M1} and~\ref{prop_M2} in Appendix~2.
In all cases the proofs are quite elementary and are based on a detailed study of a form 
(\ref{eq103x}) for~$M$.

\begin{proposition}
\label{prop_M3}
Let be $\hat b_1 = \hat b_2 = \eta_{j,l}$. 
The function $M(\eta_{j,l} , \eta_{j,l} , \hat a_1 , \hat a_2 ; \hat x)$
exists iff $a_1 = a_2$.
\end{proposition}

\noindent
{\bf Proof.}
A general Ansatz of a function $M \in \gK_1$ having a double pole
at $\eta_{j,l}$ on $\gR$ and no other poles 
is 
\[
M = \frac{1}{x- \eta_{j,l}} + c. 
\]  
(We remind that this expression corresponds to a {\em double\/} pole, since the local 
variable near $\eta_{j,l}$ is $\tau = \sqrt{x - \eta_{j,l}}$.) 
Then, $M$ has a zero at $\hat a_1$ if $c$ is chosen in such a way that 
\begin{equation}
M(\hat x) = \frac{x - a_1}{x - \eta_{j,l}}.
\label{eq014a}
\end{equation}
Indeed, the other zero on $\gR$ should have the same affix~$a_1$.
$\square$


\begin{proposition}
\label{prop_M4}
Let a double pole be located at the point $\hat b_1$, and a simple zero be located at $\hat b_2$, 
such that $b_1 = b_2 = b$ not equal to the infinity or any of the branch points. 
Let another simple zero be located at $\eta_{j,l}$. 
The function $M(\hat b_1 , \hat b_1 , \hat b_2 , \eta_{j,l} ; x)$ exists iff 
\begin{equation}
\frac{\Upsilon (\hat b_1)}{(\eta_{j,l} - b)^2} 
+
\frac{\dot \Upsilon (\hat b_1)}{\eta_{j,l} - b} 
+
\frac{\ddot \Upsilon (\hat b_1)}{2} = 0.
\label{eq014d}
\end{equation}
\end{proposition}

\noindent
{\bf Proof.}
A general 
Ansatz for a function with a double pole at $\hat b_1$ and regular at $\hat b_2$ is as follows: 
\begin{equation}
M(\hat x) = \frac{\Upsilon (\hat x)}{(x - b)^2} 
+ 
\frac{ \Upsilon (\hat b_1)}{(x - b)^2}
+ 
\frac{\dot \Upsilon (\hat b_1)}{x - b}
+
c.
\label{eq014b}
\end{equation}
Function $M(\hat x)$ has a zero at $\hat b_2$ if 
\begin{equation}
c = \frac{\ddot \Upsilon (\hat b_1)}{2}, 
\label{eq014c}
\end{equation}
where 
\[
\ddot \Upsilon (\hat x) \equiv 
\frac{d^2 \Upsilon (\hat x)}{dx^2}.
\]
Since $\Upsilon (\eta_{j,l}) = 0$, the condition $M(\eta_{j,l}) = 0$ reads as (\ref{eq014d}). 
Function $M$ is given by (\ref{eq014b}), (\ref{eq014c}). $\square$


\subsection{Building the elements $F_1$ and $F_2$ of the basis $\Omega_{3:1}$}

Here we describe the most tricky result of the paper, namely, we build functions 
$F_{1,2}$ (we remind that the choice of these functions is not unique).  

The difficulty of building $F_{1,2}$ is as follows. 
According to Theorem~\ref{th_K3}, say,
$F_1(\tilde x)$ is a cubic radical of a 
function $G_1 \in \gK_1$.
Since function $F_1$ is not allowed to have branch points
on $\gR$, all poles and zeros of 
$G(\hat x)$ on $\gR$ should have order $3 \nu$, $\nu \in \mathbb{Z}$. 
At the same time, $G_1$ cannot be a cube of another function from $\gK_1$, otherwise 
$F_1 (\Lambda(\tilde x)) = F_1 (\tilde x)$ and the condition (\ref{eq_prop1}) cannot be fulfilled.
So, it is necessary to construct a function having triple poles and zeros, but which is not a cube 
of a meromorphic function. Clearly, there are no such functions on $\overline{\mathbb{C}}$, but, surprisingly, 
such functions exist  on~$\gR$:  

\begin{theorem}
\label{th_F}
Let $\hat a$, $\hat b$, $\hat c$ be three points on $\gR$. Let $\gamma_1$, $\gamma_2$, $\gamma_3$
be oriented contours connecting the points $\hat a$ and $\hat b$, $\hat b$ and $\hat c$, 
and $\hat c$ and $\hat a$, respectively. Let the concatenation $\gamma_1 + \gamma_2 + \gamma_3$
be homotopic to the contour $\sigma_\beta$ shown in Fig.~\ref{fig22}. Let be 
\begin{equation}
\int_{\gamma_1} \frac{dx}{\Upsilon (x)} 
=
\int_{\gamma_2} \frac{dx}{\Upsilon (x)}
=
\int_{\gamma_3} \frac{dx}{\Upsilon (x)}
= 
\frac{T_\beta}{3} .
\label{eq016}
\end{equation}  
Then    
   
a) Function  
\begin{equation}
G_1 = \frac{M(\hat a , \hat a, \hat b, \hat c ; \hat x)}{M(\hat b , \hat b, \hat a, \hat c ; \hat x)}
\in \gK_1
\label{eq017}
\end{equation}
has a triple pole at $\hat a$, a triple zero 
at $\hat b$, and no other zeros or poles.    

b) Function 
\begin{equation}
G_2 = \frac{M(\hat a , \hat a, \hat b, \hat c ; \hat x)}{M(\hat c , \hat c, \hat a, \hat b ; \hat x)}
\in \gK_1
\label{eq017a}
\end{equation} 
has a triple pole at $\hat a$, a triple zero 
at $\hat c$, and no other zeros or poles. 

c) Function $F_1 = (G_1)^{1/3}$ is meromorphic on $\gR_3$ and obeys (\ref{eq_prop1}).
   
d) Function $F_2 = (G_2)^{1/3}$ is meromorphic on $\gR_3$ and obeys (\ref{eq_prop2}).
\end{theorem}

The proof of the theorem is given in Appendix~3.  
According to this proof, one can take an arbitrary point on $\gR$ as $\hat a$, find $\hat b$ and $\hat c$
as functions of $\hat a$,
and build corresponding functions $F_1$ and~$F_2$.  For convenience, 
fix the point
\begin{equation}
\hat a = \eta_{2,1}.
\label{eq_def_a}
\end{equation}
By definition (\ref{eq016}) of $\hat b$, $\hat c$, and by using the mapping $\psi$, 
one can write 
\begin{equation}
\hat b = \psi (T_\beta / 3) , \qquad \hat c = \psi (2 T_\beta / 3). 
\label{eq_hb}
\end{equation}
Introduce the notations 
\[
\hat \b = \hat b, 
\qquad
\hat \c = \hat c
\]
to stress that these $\hat b$ and $\hat c$ are some fixed values.  
For the selected  $\hat a, \hat b, \hat c$, the functions $F_{1,2}$ are given by the following proposition. 
 
\begin{proposition}
\label{prop_calc}
Let $\hat a$, $\hat b$, $\hat c$ be defined by (\ref{eq_def_a}), (\ref{eq_hb}). Then 

a) Functions $F_1(\tilde x)$ and $F_2 (\tilde x)$ have form 
\begin{equation}
F_1 (\tilde x) = \left(  \frac{\Upsilon (\hat x)}{(x-\b)^2} + 
\frac{\Upsilon (\hat \b)}{(x-\b)^2} +
\frac{\dot \Upsilon (\hat \b)}{x-\b} +
\frac{\ddot \Upsilon (\hat \b)}{2}  \right)^{-1/3} 
\left( 
\frac{
x - \b
}{
x - \eta_{2,1}
}
\right)^{1/3},
\label{eq018b}
\end{equation}
\begin{equation}
F_2 (\tilde x) = \left( - \frac{\Upsilon (\hat x)}{(x-\b)^2} + 
\frac{\Upsilon (\hat \b)}{(x-\b)^2} +
\frac{\dot \Upsilon (\hat \b)}{x-\b} +
\frac{\ddot \Upsilon (\hat \b)}{2}  \right)^{-1/3} 
\left( 
\frac{
x - \b
}{
x - \eta_{2,1}
}
\right)^{1/3}.
\label{eqF2}
\end{equation}

b) The affix $\b$ of $\hat b$ obeys a fourth-order algebraic equation 
\begin{equation}
h_0 + h_1 \b+ h_2 \b^2 +h_3 \b^3+h_4 \b^4 = 0,
\label{eq019a}
\end{equation} 
where
\[
h_0 =  \eta_{2,1} + 3\eta_{1,2} - \eta_{2,1}^2\eta_{1,2} + \eta_{2,1}\eta_{1,2}^2,
\]
\[
h_1 = -4(1+2\eta_{2,1}\eta_{1,2}+\eta_{1,2}^2),
\]
\[
h_2 = 6(\eta_{2,1}+\eta_{1,2} + \eta_{2,1}^2\eta_{1,2}+\eta_{2,1}\eta_{1,2}^2),
\]
\[
h_3 = -4\eta_{2,1}(\eta_{2,1}+2\eta_{1,2} + \eta_{2,1}\eta_{1,2}^2),
\] 
\[
h_4 = \eta_{2,1}-\eta_{1,2} + 3\eta_{2,1}^2\eta_{1,2} + \eta_{2,1}\eta_{1,2}^2.
\] 

c) The affix $\b$ can be found by solving
the ordinary differential equation 
\begin{equation}
\frac{d x}{d \chi} = \Upsilon (x)
\label{eq022}
\end{equation}
for the function $x (\chi)$
on the segment $\chi \in [0 , T_\beta / 3]$. The initial data is $x(0) = \eta_{2,1}$. 
The result is defined by  $\b = x (T_\beta / 3)$.
\end{proposition}

\noindent  
{\bf Proof.}  a)
Construct the ratios (\ref{eq017}), (\ref{eq017a}) for the selected points
$\hat a$, $\hat b$, $\hat c$.
 Note that according Proposition~\ref{prop_M3}, 
the affixes of $\hat b$ and $\hat c$ coincide, and they are located on different sheets of $\gR$:
\[
\b = \c, \qquad \hat \c = \Pi(\hat \b).
\]
Besides, this proposition yields
\begin{equation}
M(\eta_{2,1} , \eta_{2,1} , \hat \b , \hat \c ; \hat x) = \frac{x - \b}{x - \eta_{2,1}}. 
\label{eq_prf1}
\end{equation}
According to Proposition~\ref{prop_M4}, 
\begin{equation}
M(\hat \b , \hat \b , \eta_{2,1} , \hat \c ; \hat x) = 
\frac{\Upsilon (\hat x)}{(x-\b)^2} + 
\frac{\Upsilon (\hat \b)}{(x-\b)^2} + 
\frac{\dot \Upsilon (\hat \b)}{x-\b} + 
\frac{\ddot \Upsilon (\hat \b)}{2} 
\label{eq_prf2}
\end{equation}
provided 
\begin{equation}
\frac{\Upsilon (\hat \b)}{(\eta_{2,1}-\b)^2} + 
\frac{\dot \Upsilon (\hat \b)}{\eta_{2,1}-\b} + 
\frac{\ddot \Upsilon (\hat \b)}{2} =0 
\label{eq019}
\end{equation}
is valid.  According to the same proposition, 
\begin{equation}
M(\hat \c , \hat \c , \eta_{2,1} , \hat \b ; \hat x) = - 
\frac{\Upsilon (\hat x)}{(x-\b)^2} + 
\frac{\Upsilon (\hat \b)}{(x-\b)^2} + 
\frac{\dot \Upsilon (\hat \b)}{x-\b} + 
\frac{\ddot \Upsilon (\hat \b)}{2}. 
\label{eq_prf4}
\end{equation}
(we change the sign of (\ref{eq014b}) for convenience; it can be done since all $M$-functions are defined up to 
a constant factor). The condition of existence of $M(\hat \c , \hat \c , \eta_{2,1} , \hat \b ; \hat x)$
is also (\ref{eq019}). Substituting (\ref{eq_prf1}), (\ref{eq_prf2}), and (\ref{eq_prf4}) into 
(\ref{eq017}) and (\ref{eq017a}), obtain (\ref{eq018b}) and (\ref{eqF2}).

Let us prove point b). Consider the condition (\ref{eq019}). Divide it by 
$\Upsilon(\hat \b)$ and note that ratios $\dot \Upsilon (\hat \b) / \Upsilon (\hat \b)$
and $\ddot \Upsilon(\hat \b) / \Upsilon(\hat \b)$ are rational functions of $\b$. Thus, 
(\ref{eq019}) is an algebraic equation for $\b$. 
After some algebra get (\ref{eq019a}). 

The proof of c) is elementary: (\ref{eq022}) is another form of (\ref{eq005}). $\square$

\vskip 6pt

Several remarks should be made regarding Proposition~\ref{prop_calc}.   

\vskip 6pt
\noindent
{\bf Remark 1. }
To solve the ordinary differential equation (\ref{eq022}) near 
the branch point $\eta_{2,1}$ one can use the local variable
on $\gR$, namely,  
\[
\tau = \tau(x) = \sqrt{x - \eta_{2,1}},
\] 
as the dependent variable for the ODE. 
One can rewrite  (\ref{eq022}) as  
\begin{equation}
\frac{d \tau}{d \chi} 
=
\frac{1}{2}
\sqrt{
(\tau^2 + \eta_{2,1} - \eta_{1,1})
(\tau^2 + \eta_{2,1} - \eta_{1,2})
(\tau^2 + \eta_{2,1} - \eta_{2,2})
}
. 
\label{eq_ODE} 
\end{equation}
Thus, one can solve (\ref{eq_ODE}) in some small neighborhood of $\eta_{2,1}$, 
and then solve (\ref{eq022}) on the remaining part of the segment $\chi \in [0 , T_\beta/ 3]$. 
 
\vskip 6pt 
\noindent
{\bf Remark 2.} 
By construction, $F_1$ has three simple poles at $P_{3:1}^{-1} (\eta_{2,1})$ and
three simple zeros at $P_{3:1}^{-1} (\hat \b)$. There are no other poles or zeros. 
The properties of $F_2$ are similar, but it has simple zeros at $P_{3:1}^{-1} (\hat \c)$. 
 
\vskip 6pt
\noindent
{\bf Remark 3.}  
The algebraic condition (\ref{eq019a}) is slightly weaker than the transcendent condition
(\ref{eq017a}). Namely, equation (\ref{eq019a}) is fulfilled
if there exist {\em any\/} contours $\gamma_1$, $\gamma_2$, $\gamma_3$
cyclically connecting the 
points $\eta_{2,1}$ and two points on $\gR$ having affix $\b$, such that 
\begin{equation}
\int_{\gamma_1} \frac{dx}{\Upsilon (\hat x)}  
=
\int_{\gamma_2} \frac{dx}{\Upsilon (\hat x)}
=
\int_{\gamma_3} \frac{dx}{\Upsilon (\hat x)}.
\label{eq019b}
\end{equation}   
The concatenation of the contours $\gamma_1 + \gamma_2 + \gamma_3$ is not 
necessarily homotopic to~$\sigma_\beta$, thus, each of the integrals is not necessarily 
equal to $T_\beta /3$.   

A detailed study shows that 
the equation (\ref{eq019a})
has four roots: 
$b_1$, $b_2$, $b_3$, $b_4$,
such that 
\begin{equation}
\int \limits_{\eta_{2,1}}^{b_1} \frac{dx}{\Upsilon (\hat x)}
= \pm \frac{T_\beta}{3} + \mu T_\beta + \nu T_\alpha, 
\label{eq020a}
\end{equation}
\begin{equation}
\int \limits_{\eta_{2,1}}^{b_2} \frac{dx}{\Upsilon (\hat x)}
= \pm \frac{T_\alpha}{3} + \mu T_\beta + \nu T_\alpha, 
\label{eq020b}
\end{equation}\begin{equation}
\int \limits_{\eta_{2,1}}^{b_3} \frac{dx}{\Upsilon (\hat x)}
= \pm \frac{T_\alpha + T_\beta}{3} + \mu T_\beta + \nu T_\alpha, 
\label{eq020c}
\end{equation}\begin{equation}
\int \limits_{\eta_{2,1}}^{b_4} \frac{dx}{\Upsilon (\hat x)}
= \pm \frac{T_\alpha - T_\beta}{3} + \mu T_\beta + \nu  T_\alpha. 
\label{eq020d}
\end{equation}
The integrals are defined up to the sign and up to the integers $\mu$, $\nu$, 
which depend on the particular 
choice of the integration contour. 
One can see that only $b_1$ fits the condition (\ref{eq017a}), i.~e.\ $\b = b_1$.

\vskip 6pt
\noindent
{\bf Remark 4.} 
The functions $F_{1,2}$ are defined by (\ref{eq018b}) ambiguously. 
This ambiguity follows from that of the cubic radical, i.~e.\ the result can be 
multiplied by $\varpi$ or $\varpi^{-1}$. 
Let us remove this ambiguity. 
The symmetry $\Pi$ converts $F_1$ into $F_2$ (up to multiplication by some cubic root of~1):
\begin{equation}
F_1 (\Pi(\tilde x)) = \delta F_2 (\tilde x), 
\qquad \delta \in \{ 1, \varpi , \varpi^{-1} \}. 
\label{eqF2_1}
\end{equation} 
To prove this, use (\ref{eq306a}) and note that 
\[
\frac{\Upsilon (\hat x)}{(x-\b)^2} + 
\frac{\Upsilon (\hat \b)}{(x-\b)^2} +
\frac{\dot \Upsilon (\hat \b)}{x-\b} +
\frac{\ddot \Upsilon (\hat \b)}{2}
\stackrel{\Pi}{\longrightarrow}
-\frac{\Upsilon (\hat x)}{(x-\b)^2} + 
\frac{\Upsilon (\hat \b)}{(x-\b)^2} +
\frac{\dot \Upsilon (\hat \b)}{x-\b} +
\frac{\ddot \Upsilon (\hat \b)}{2} 
\]
To remove some of the ambiguity of determining $F_1$ and $F_2$, 
fix the value $\delta = 1$, thus fixing 
\begin{equation}
F_1 (\Pi(\tilde x)) = F_2 (\tilde x). 
\label{eqF2_2}
\end{equation} 
Thus, one can choose the cubic root defining $F_1$ arbitrarily, and then the choice 
for $F_2$ follows from (\ref{eqF2_2}).

\vskip 6pt
\noindent
{\bf Remark 5.} 
The product of functions $F_1$ and $F_2$ is rational. Namely, it belongs 
to $\gK_1$ since
\[
F_1 (\Lambda(\tilde x)) F_2 (\Lambda(\tilde x)) 
=
F_1 (\tilde x) F_2 ( \tilde x ),
\]
and, then, it belongs to $\gK_0$ since
\[
F_1 (\Pi(\tilde x)) F_2 (\Pi(\tilde x)) 
=
F_1 (\tilde x) F_2 ( \tilde x ).
\]
One can easily see that $F_1 (\tilde x) F_2(\tilde x)$
should have a simple pole at $x = \eta_{2,1}$ and a simple zero at 
$x = \b$. Studying the function at infinity, one can find that 
\begin{equation}
F_1 (\tilde x) F_2 ( \tilde x ) = 
\frac{1}{((\ddot \Upsilon (\hat \b))^2 / 4-1)^{1/3}}
 \frac{x - \b}{x - \eta_{2,1}}. 
\label{eqF2_3}
\end{equation}



\section{Constructing the Sommerfeld transformant $A(x)$}
\label{sec_span}

Here we assume that $x_{\rm in}$ and $x_{\rm in}^{-1}$ are not
equal to $\b$ or to $\eta_{j,l}$. 

Let us build the Sommerfeld transformant $A(\tilde x)$ 
in the form (\ref{eq306m}), i.~e.\ let us find functions 
$A_0, A_1, A_2$ obeying Functional problem~2.
These functions have form (\ref{eq107a}),    
where 
the coefficients $q_j (\hat x)$ belong to~$\gK_1$. 

According to Proposition~\ref{prop_K1},  
\begin{equation}
q_j (\hat x) = q_j' (x) + q_j'' (x) \Upsilon (\hat x),
\qquad 
j = 0,1,2,
\label{eq601}
\end{equation}  
where 
$q_j' (x)$ and $q_j'' (x)$ belong to $\gK_0$, i.~e.\ they are rational functions of~$x$. 
The aim of this section is to find the functions 
$q_j' (x)$ and~$q_j'' (x)$.

Functions $A_0,A_1,A_2$ obey conditions a) and~d) of Functional problem~2 by construction. 
Condition~b) and~c) are provided by Propositions~\ref{prop_coefs1} and~\ref{prop_coefs2}, 
respectively.    

\begin{proposition}
\label{prop_coefs1}
Functions $A_{0,1,2}$ defined by (\ref{eq107a}) obey condition~b) of Functional problem~2 iff 
the coefficients
$q_j'(x)$, $q_j''(x)$ 
have form  
\begin{equation}
q_0' (x) = \frac{s_1}{x - x_{\rm in}} + \frac{s_2}{x - x_{\rm in}^{-1}} + s_0 , 
\label{eq602}
\end{equation}
\begin{equation}
q_0'' (x) = \frac{s_3}{x - x_{\rm in}} -  \frac{s_3}{x - x_{\rm in}^{-1}},
\label{eq603}
\end{equation}
\begin{equation}
q_1'  (x) = (x - \eta_{2,1}) \left( 
\frac{s_4 }{x - x_{\rm in}}
+ 
\frac{s_5 }{x - x_{\rm in}^{-1}}
- 
\frac{(s_6+s_7) \Upsilon (\hat \b)}{x - \b}
\right) ,
\label{eq604}
\end{equation}
\begin{equation}
q_1'' (x) = 
(\b - \eta_{2,1}) \left( 
\frac{s_6}{x - x_{\rm in}}
+ 
\frac{s_7}{x - x_{\rm in}^{-1}}
- 
\frac{s_6 + s_7 }{x - \b} \right),
\label{eq605}
\end{equation}
\begin{equation}
q_2'  (x) = (x - \eta_{2,1}) \left( 
\frac{s_8 }{x - x_{\rm in}}
+ 
\frac{s_9 }{x - x_{\rm in}^{-1}}
+
\frac{(s_{10}+s_{11}) \Upsilon (\hat \b)}{x - \b}
\right) ,
\label{eq606}
\end{equation}
\begin{equation}
q_2'' (x) = 
(\b - \eta_{2,1}) \left( 
\frac{s_{10}}{x - x_{\rm in}}
+ 
\frac{s_{11}}{x - x_{\rm in}^{-1}}
- 
\frac{s_{10} + s_{11} }{x - \b} \right).
\label{eq607}
\end{equation}
where $s_0, \dots , s_{11}$ are some arbitrary parameters. 
\end{proposition} 

\noindent
{\bf Proof. }
To find the Ansatz for each of the coefficients $q'_j$, $q''_j$ let us prove the following statements based on Functional 
problem~2.

\vskip 6pt
\noindent 
{\bf 1. } {\em Let $x$ be not equal to $\infty$, $\eta_{j,l}$, $\b$, $x_{\rm in}$, or 
$x_{\rm in}^{-1}$. 
Then all functions $q_j' $, $q_j'' $, $j = 0,1,2$ are regular at~$x$.}

The proof is as follows. Consider some particular~$j$.
Let $\nu$ be the highest pole order of the functions $q_j' $, $q_j'' $
at~$x$. 
Note that  $F_{1,2} (x) \ne 0$ and $\Upsilon (x) \ne 0$. Thus, the pole of order  
$\nu$ will appear on the sheet~1 or~2 (corresponding residues cannot be compensated both). This 
contradicts to Functional problem~2. 

The statements~2, 3, and~4 are similar to  statement~1, so we omit their proofs.

\vskip 6pt
\noindent
{\bf 2.} {\em The functions $q_k'$, $q''_k$, $k = 0,1,2$ are regular at the points 
$\eta_{1,1}$, $\eta_{1,2}$, $\eta_{2,2}$.}  

\vskip 6pt
\noindent
{\bf 3.} {\em The functions $q_k'$, $x^2 q''_k$, $k = 0,1,2$ are regular at infinity.}  

\vskip 6pt
\noindent
{\bf 4.} {\em The functions $q_k'$, $q''_k$ have simple poles 
at $x_{\rm in}$ and $x_{\rm in}^{-1}$. }

Slightly more subtle consideration is needed for the values $x$ equal to 
$\eta_{2,1}$ and $\b$, since functions $F_1$ and $F_2$ have poles and zeros at these affixes. 
The following statements can be checked:

\vskip 6pt
\noindent
{\bf 5.} {\em  The functions 
$q_0'(x)$, $q_0''(x)$,
$(x - \eta_{2,1})^{-1} q_1'(x)$, $q_1''(x)$,
$(x - \eta_{2,1})^{-1} q_2'(x)$, $q_2''(x)$ 
are regular at $x = \eta_{2,1}$.} 

\vskip 6pt 
\noindent
{\bf 6.} {\em Functions $q_0'$ and $q_0''$ are regular at $\b$.  
$q_j' (x)$, $q_j'' (x)$, $j = 1,2$ can have simple poles at $\b$. 
The following identities should be valid: 
\[
\lim_{x \to \b}[q_1'(x) - \Upsilon (\hat \b) q_1''(x)] = 0, 
\qquad 
\lim_{x \to \b}[q_2'(x) + \Upsilon (\hat \b) q_2''(x)] = 0, 
\]}

One can see that (\ref{eq602})--(\ref{eq607}) are the most general formulae for rational 
functions obeying statements 1--6.  
$\square$

\begin{proposition}
\label{prop_coefs2}
Functions $A_{0,1,2}$ defined by (\ref{eq107a}) obey condition~c) of Functional problem~2 iff
Proposition~\ref{prop_coefs1} is fulfilled, and 

\begin{equation}
s_1 = i Y_1 / (6 \pi) ,
\qquad 
s_2 = i Y_3 / (6 \pi) , 
\qquad 
s_3 = 0,
\label{eq608}
\end{equation}

\begin{equation}
(x_{\rm in} - \eta_{2,1}) s_4 F_1(\tilde x_1) + (\b - \eta_{2,1}) s_6 Y_1 F_1 (\tilde x_1) 
= i Y_1  / (6 \pi), 
\label{eq609}
\end{equation}
\begin{equation}
(x_{\rm in} - \eta_{2,1}) s_4 F_1(\tilde x_2) - (\b - \eta_{2,1}) s_6 Y_1 F_1 (\tilde x_2) 
= i Y_1 / (6 \pi), 
\label{eq610}
\end{equation} 
 
\begin{equation}
(x_{\rm in} - \eta_{2,1}) s_8 F_2(\tilde x_1) + (\b - \eta_{2,1}) s_{10} Y_1 F_2 (\tilde x_1) 
= i Y_1 / (6 \pi), 
\label{eq611}
\end{equation}
\begin{equation}
(x_{\rm in} - \eta_{2,1}) s_8 F_2(\tilde x_2) - (\b - \eta_{2,1}) s_{10} Y_1 F_2 (\tilde x_2) 
= i Y_1 / (6 \pi), 
\label{eq612}
\end{equation} 
 
\begin{equation}
(x_{\rm in}^{-1} - \eta_{2,1}) s_5 F_1(\tilde x_3) + (\b - \eta_{2,1}) s_7 Y_3 F_1 (\tilde x_3) 
= i Y_3  / (6 \pi), 
\label{eq613}
\end{equation}
\begin{equation}
(x_{\rm in}^{-1} - \eta_{2,1}) s_5 F_1(\tilde x_4) - (\b - \eta_{2,1}) s_7 Y_3 F_1 (\tilde x_4) 
= i Y_3 / (6 \pi), 
\label{eq614}
\end{equation} 

\begin{equation}
(x_{\rm in}^{-1} - \eta_{2,1}) s_9 F_2(\tilde x_3) + (\b - \eta_{2,1}) s_{11} Y_3 F_2 (\tilde x_3) 
= i Y_3  / (6 \pi), 
\label{eq615}
\end{equation}
\begin{equation}
(x_{\rm in}^{-1} - \eta_{2,1}) s_9 F_2(\tilde x_4) - (\b - \eta_{2,1}) s_{11} Y_3 F_2 (\tilde x_4) 
= i Y_3 / (6 \pi). 
\label{eq616}
\end{equation} 

The additive parameter $s_0$ can be chosen arbitrarily (e.g.\ $s_0 = 0$).
\end{proposition}

The proof is straightforward. One should substitute (\ref{eq602})--(\ref{eq607}) into (\ref{eq601}), (\ref{eq107a})
and check the validity of condition~c) of Functional problem~2. 

The equations (\ref{eq609})--(\ref{eq616}) can be easily solved. The result can be written using 
(\ref{eqF2_3}) as follows: 

\begin{equation}
s_4 = Z \Upsilon (\hat x_1) 
\frac{1}{x_{\rm in} - \b} (F_2(\tilde x_1) + F_2 (\tilde x_2)),
\label{eq617}
\end{equation} 
\begin{equation}
s_6 = Z 
\left( 
\frac{1}{ \b - \eta_{2,1}}
+
\frac{1}{x_{\rm in} - \b}
\right) 
(F_2(\tilde x_1) - F_2 (\tilde x_2)),
\label{eq618}
\end{equation} 

\begin{equation}
s_8 = Z \Upsilon (\hat x_1)
\frac{1}{x_{\rm in} - \b} (F_1(\tilde x_1) + F_1 (\tilde x_2)),
\label{eq619}
\end{equation} 
\begin{equation}
s_{10} = Z 
\left( 
\frac{1}{ \b - \eta_{2,1}}
+
\frac{1}{x_{\rm in} - \b}
\right) 
(F_1(\tilde x_1) - F_1 (\tilde x_2)),
\label{eq620}
\end{equation} 

\begin{equation}
s_5 = Z \Upsilon (\hat x_3)
\frac{1}{x_{\rm in}^{-1} - \b} (F_2(\tilde x_3) + F_2 (\tilde x_4)),
\label{eq621}
\end{equation} 
\begin{equation}
s_7 = Z 
\left( 
\frac{1}{ \b - \eta_{2,1}}
+
\frac{1}{x_{\rm in}^{-1} - \b}
\right) 
(F_2(\tilde x_3) - F_2 (\tilde x_4)),
\label{eq622}
\end{equation} 

\begin{equation}
s_9 =Z \Upsilon (\hat x_3)
\frac{1}{x_{\rm in}^{-1} - \b} (F_1(\tilde x_3) + F_1 (\tilde x_4)),
\label{eq623}
\end{equation} 
\begin{equation}
s_{11} = Z 
\left( 
\frac{1}{ \b - \eta_{2,1}}
+
\frac{1}{x_{\rm in}^{-1} - \b}
\right) 
(F_1(\tilde x_3) - F_1 (\tilde x_4)),
\label{eq624}
\end{equation} 
where 
\begin{equation}
Z =\frac{i ((\ddot \Upsilon (\hat \b))^2 / 4 - 1)^{1/3}}{12 \pi}.
\label{eq624a}
\end{equation}

Finally, the Sommerfeld transformant is found. 
The formulae that should be used for computations are 
(\ref{eq617})--(\ref{eq624}),
(\ref{eq602})--(\ref{eq607}),
(\ref{eq107}),
(\ref{eq018b}),
(\ref{eqF2}). 
The value $\b$ is defined from (\ref{eq019a}) or from the ODE (\ref{eq022}). 
The transformant is substituted into the Sommerfeld integral (\ref{eq212}).

\section{Numerical examples}
\label{sec_num}

In this section we are demonstrating the ideas of the paper using some numerical
examples. We take real values of $K$, 
having in mind the limit 
${\rm Im}[K] \to +0$.

\subsection{Computation of  $\hat \b$}
\label{subsec_cb}

For practical computations, we propose two following algorithms for 
computing the value $\b$ with a high accuracy.

\vskip 6pt

\noindent
{\bf Algorithm~1:} 
\nopagebreak

\begin{enumerate}

\item
Compute $T_\beta$ by numerical integration. 

\item
Find $\b$ approximately by solving numerically the ordinary differential equation 
(\ref{eq022})
on the segment $\chi \in [0 , T_\beta / 3]$. Take $x(0) = \eta_{2,1}$. 
The value $x (T_\beta / 3)$ is the approximation for~$\b$.
Denote it by~$\b'$.

\item
Using $\b'$ as a starting approximation, solve (\ref{eq019a}) by Newton's method. 
As a result, after several iterations, get a refined value of~$\b$ with a machine accuracy.

\end{enumerate} 

Since the first two steps are necessary only to obtain the starting approximation 
for Newton's method used on the third step, very coarse meshes can be used 
for numerical integration and for solving the ordinary differential equation.
The Newton's method is very cheap, and, several iterations provide the value of $\b$ 
having the machine accuracy. 

Another algorithm can be developed, taking the algebraic equation
(\ref{eq019a}) as the starting point. The algorithm is as follows. 

\vskip 6pt

\noindent
{\bf Algorithm~2:} 
\nopagebreak
\begin{enumerate}

\item
Solve equation (\ref{eq019a}) and find four values: $b_1$, $b_2$, $b_3$, $b_4$. 
This can be done even explicitly. 

\item
For each affix $b_j$ take points $\hat b_j$ and $\hat b'_j$ on two different  sheets 
of $\gR$. Totally there will be eight candidates for $\hat \b$.

\item
For each value $\hat b_j$ or $\hat b_j'$ construct the function 
\begin{equation}
G_1 (\tilde x) = \left(  \frac{\Upsilon (\hat x)}{(x- b)^2} + 
\frac{\Upsilon (\hat b)}{(x-b)^2} +
\frac{\dot \Upsilon (\hat b)}{x-b} +
\frac{\ddot \Upsilon (\hat b)}{2}  \right)^{-1} 
\frac{
x - b
}{
x - \eta_{2,1}
}
,
\label{eq018bb}
\end{equation}
and check the variation of ${\rm Arg}[G_1]$ along the contour $\sigma_\beta$.
and $\sigma_\alpha$.
There should exist only one value of $\hat b$ (among the eight candidates),
for which the conditions (\ref{eq_var1}) and (\ref{eq_var2}) are valid. 
This value of $\hat \b$ is what we are looking for.  

\end{enumerate}
 
Indeed, Algorithm~1 and Algorithm~2 should yield the same result. 
We use Algorithm~1 below. 

\begin{figure}[ht]
\centering\includegraphics[width = 6.18cm]{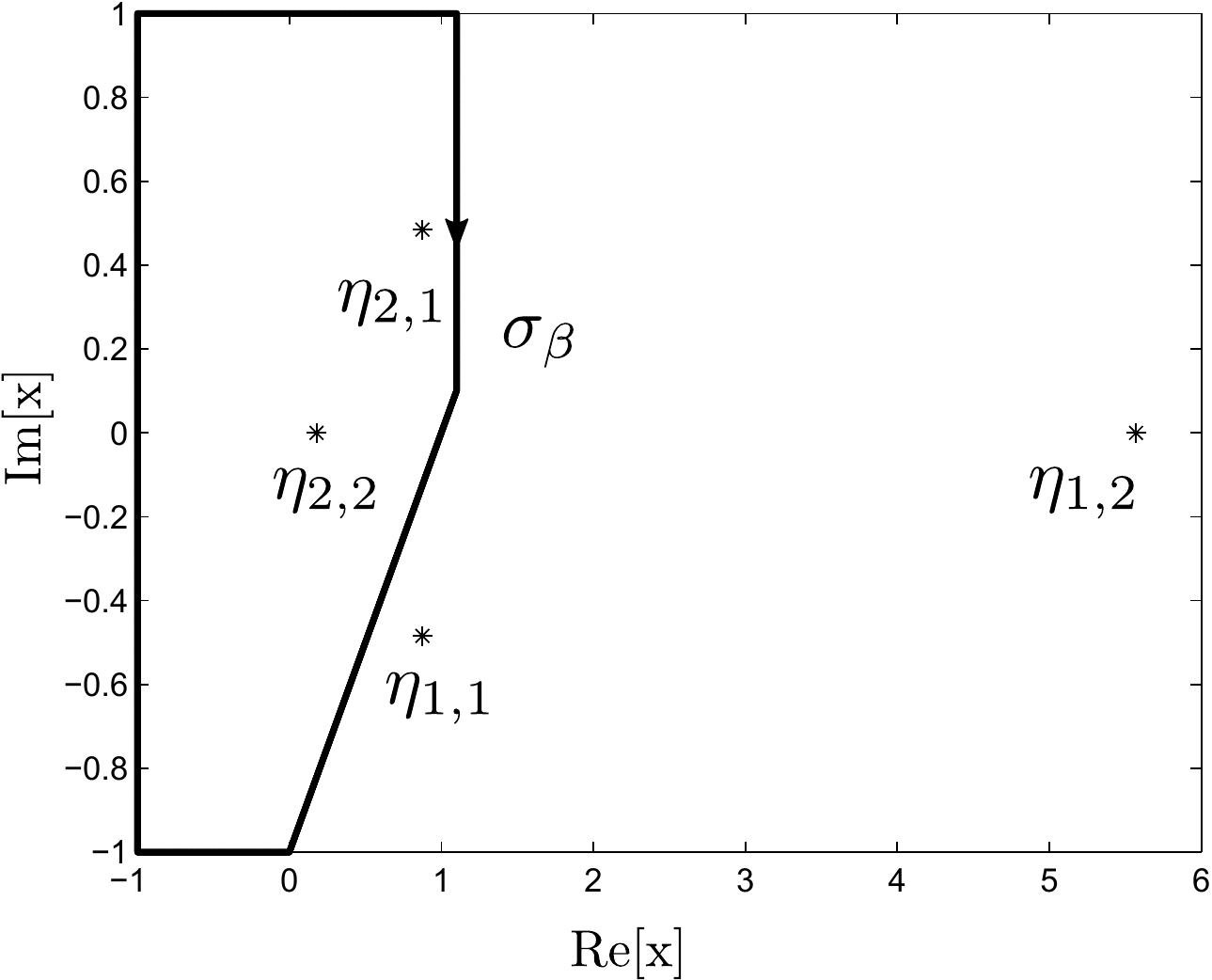}
\includegraphics[width = 7cm]{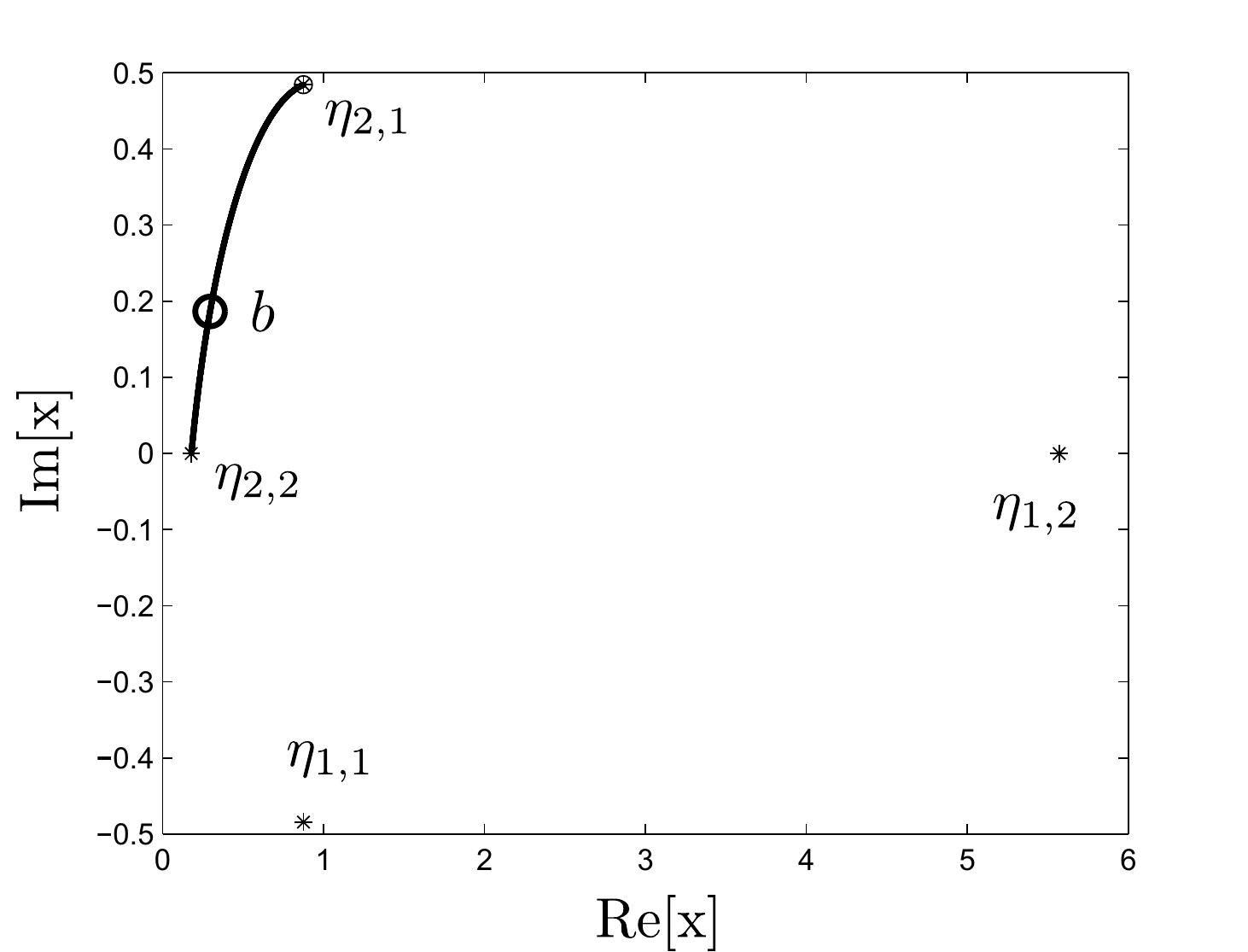}
\caption{Contour for finding $T_\beta$ (left),
solution of equation (\ref{eq022}), right}
\label{num_cont}
\end{figure}

Let be $K = 0.5$.  
First, find the period $T_\beta$ (see (\ref{eq005b})).
For the numerical integration, use contour $\sigma_\beta$ shown in 
Fig.~\ref{num_cont},~left. The positions of the branch points $\eta_{j,l}$
are shown by stars.

Find the correct values of $\Upsilon(\hat x)$ on this contour.
The contour passes through the point $x = 1$ on sheet~1, thus, one can fix 
$\Upsilon (1)$, and then utilize the continuity.
One can see that $\Upsilon (1) = \pm 0.9682i$, and one should choose the correct sign. 
Take $K$ close to 0.5, but having a small positive imaginary part, say $K = 0.5 + 0.1i$. 
The  values of $\Upsilon(1)$ for this $K$
are $\pm (-0.1810 + 0.9722i)$,
and they correspond to the values $y_1 = 0.7895 + 0.4361$ and 
$y_2 = 0.9705 - 0.5361$, respectively. One can see that $|y_1| < 1$ and $|y_2| > 1$.
Thus, one should choose  $\Upsilon(1)  = -0.1810 + 0.9722i$ for $K = 0.5 + 0.1i$. 
By continuity, $\Upsilon(1)  = 0.9682i$ for $K = 0.5$. 
This reasoning yields also that ${\rm Im}[\Upsilon (1)] > 0$ 
on the physical sheet for all real $0<K <2$.

The integral for $T_\beta$ can be easily computed for $K = 0.5$:
\[
T_\beta = -1.6219 + 2.4884i. 
\]

For demonstration purposes, solve the equation (\ref{eq022})
(or its equivalent form (\ref{eq_ODE}))
on the segment $\chi \in [0, T_\beta]$ taking the initial value
$x(0) = \eta_{2,1}$.  
The result is the trajectory going from $\eta_{2,1}$ to $\eta_{2,2}$
along one of the sheets of $\gR$ and returning back along another sheet.
The trajectory ends almost exactly at $\eta_{2,1}$. 
The projection of this trajectory onto the $x$-plane is shown in 
Fig.~\ref{num_cont},~right.

According to Algorithm~1, solve equation (\ref{eq022}) for
$\chi \in [0, T_\beta / 3]$, with
$x(0) = \eta_{2,1}$. As the result, get the position of 
the point $\hat \b$, i.~e.\ the affix $\b$ and the value 
$\Upsilon (\hat \b)$. We use this value as a starting approximation 
for~$\b$ and refer to it as $\b'$. If the ODE is solved by the simplest Euler's 
scheme on a mesh of 100 nodes, 
\[
\b'= 0.2917 + 0.1858i,
\qquad 
\Upsilon (\hat \b') =  -0.2437 + 0.7958i.
\] 
The position of $\b'$ is shown in Fig.~\ref{num_cont}, right,
by a circle. The value of $\Upsilon (\hat \b')$ is needed to conclude 
that $\hat \b'$ belongs to sheet~2 of~$\gR$. 

The value of $\b'$ obtained so far can be considered as a rough approximation 
for this parameter. According to Algorithm~1, one can solve (\ref{eq019a})
by the Newton's method to refine the value. 
The process stabilizes after 4 steps, and the result
is: 
\[
\b = 0.295390040273516 + 0.186354378894278i.
\]
One can see that the starting approximation $\b'$ happens to be quite close to the
exact root of (\ref{eq019a}). This is a clear demonstration of consistency
of our approach.   
 
\begin{figure}[ht]
\centering
\includegraphics[width = 12cm]{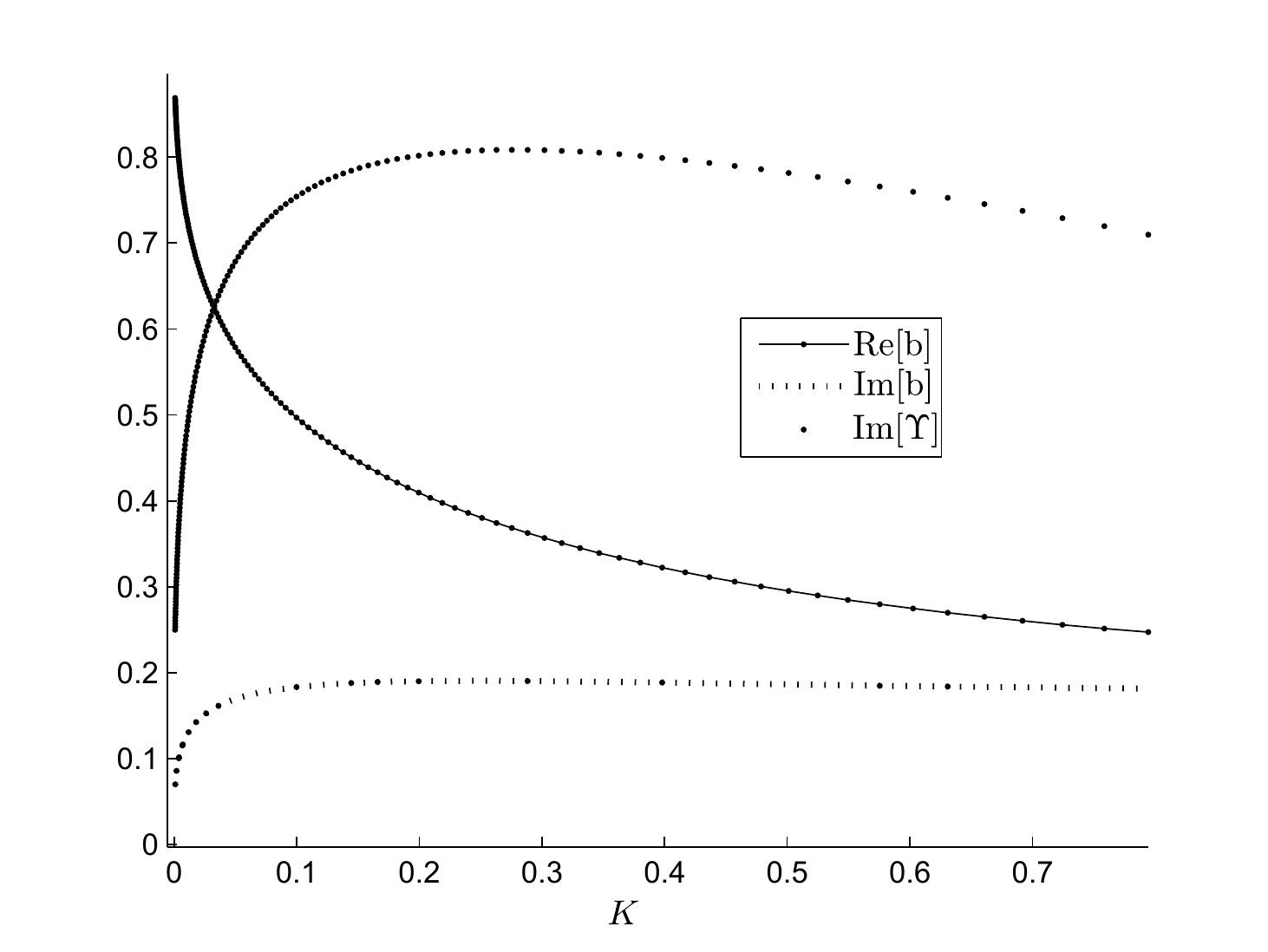}
\caption{The values of ${\rm Re}[\b]$, ${\rm Im}[\b]$,
and ${\rm Im}[\Upsilon(\hat \b)]$ as functions of $K$}
\label{numeric_b_graph}
\end{figure}

Taking $K$ belonging to a dense grid covering the segment  $[10^{-3}, 1]$
and repeating the procedure described above, 
one can obtain the values $\hat \b (K)$. They are presented graphically 
in Fig.~\ref{numeric_b_graph}. 
The affix $\b$ is shown by its real and imaginary part. 
The value $\Upsilon (\hat \b)$ is necessary only to select a correct sheet of $\gR$,
so we displayed the imaginary part of it. 

An important conclusion that can be made from Fig.~\ref{numeric_b_graph}
(and of course one can prove this analytically) is that 
\[
\b \to 1 \quad  \mbox{as} \quad K \to 0, 
\] 
and $\hat \b$ is located on sheet~2.  

\subsection{Examination of $G_1 = (F_1)^3$}

Fix $K = 0.5$ and use the value of $\hat \b$ found in the previous subsection. 
Construct the function $G_1 (\hat x)$ by the formula
(\ref{eq018bb}).
Check numerically the validity of the conditions (\ref{eq_var1}), (\ref{eq_var2}).

To check this, we build hodographs of $G_1$ on $\sigma_\alpha$ and on $\sigma_\beta$, 
i.~e.\ we plot the values of $G_1(\hat x)$ for $\hat x$ running along 
the contours $\sigma_\alpha$ and $\sigma_\beta$. As the result, we get 
oriented contours in the complex plane of~$G_1$.

\begin{figure}[ht]
\centering
\includegraphics[width = 8cm]{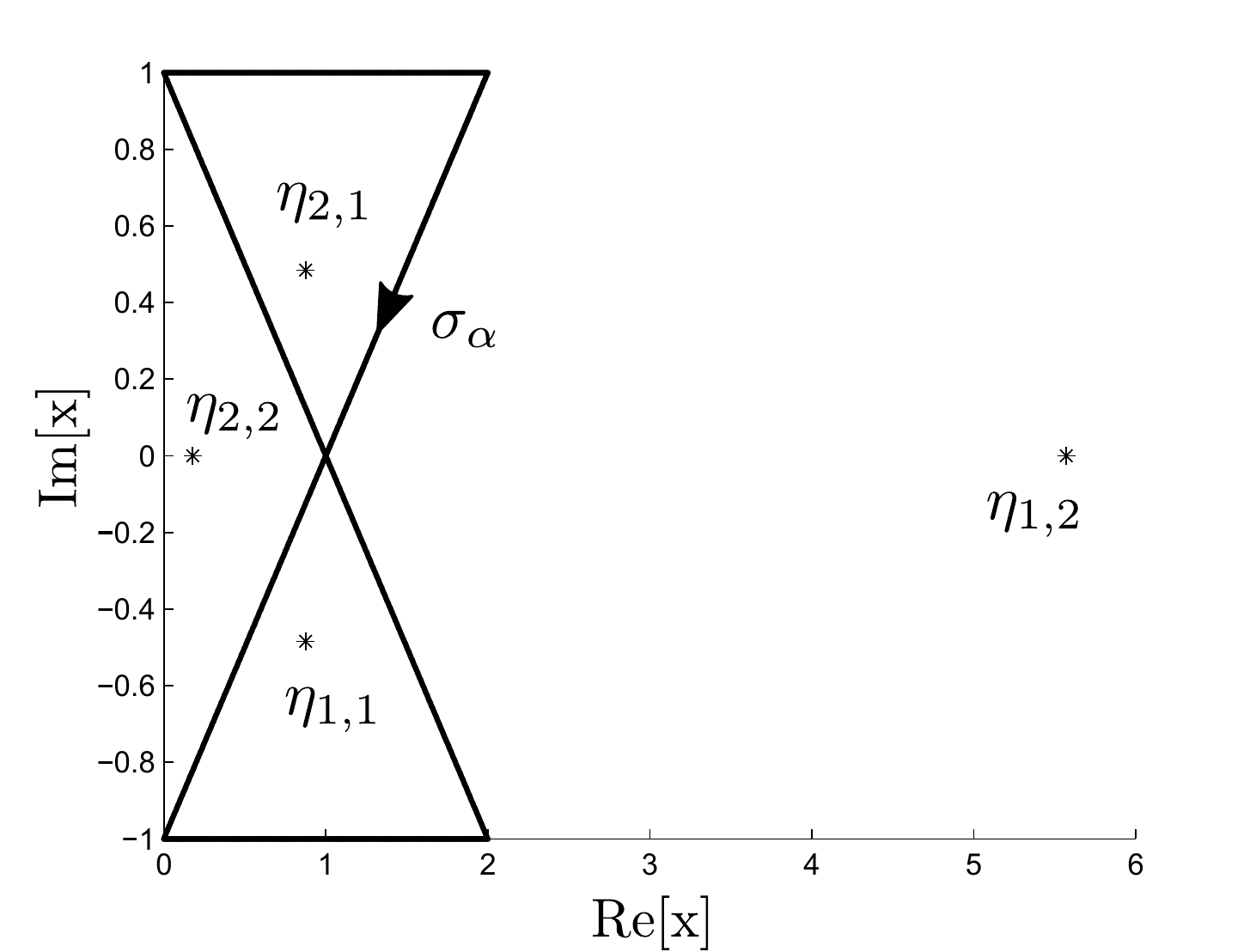}
\caption{A contour homotopic to $\sigma_\alpha$}
\label{num_sig_a}
\end{figure}

The contour homotopic to $\sigma_\beta$ has been already built (see Fig.~\ref{num_cont},~left).
The contour homotopic to $\sigma_\alpha$ and convenient for numerical 
computations is shown in Fig.~\ref{num_sig_a}. 
The contour passes the value $x = 1$ on sheet~1 on the way down. 

\begin{figure}[ht]
\centering
\includegraphics[width = 8cm]{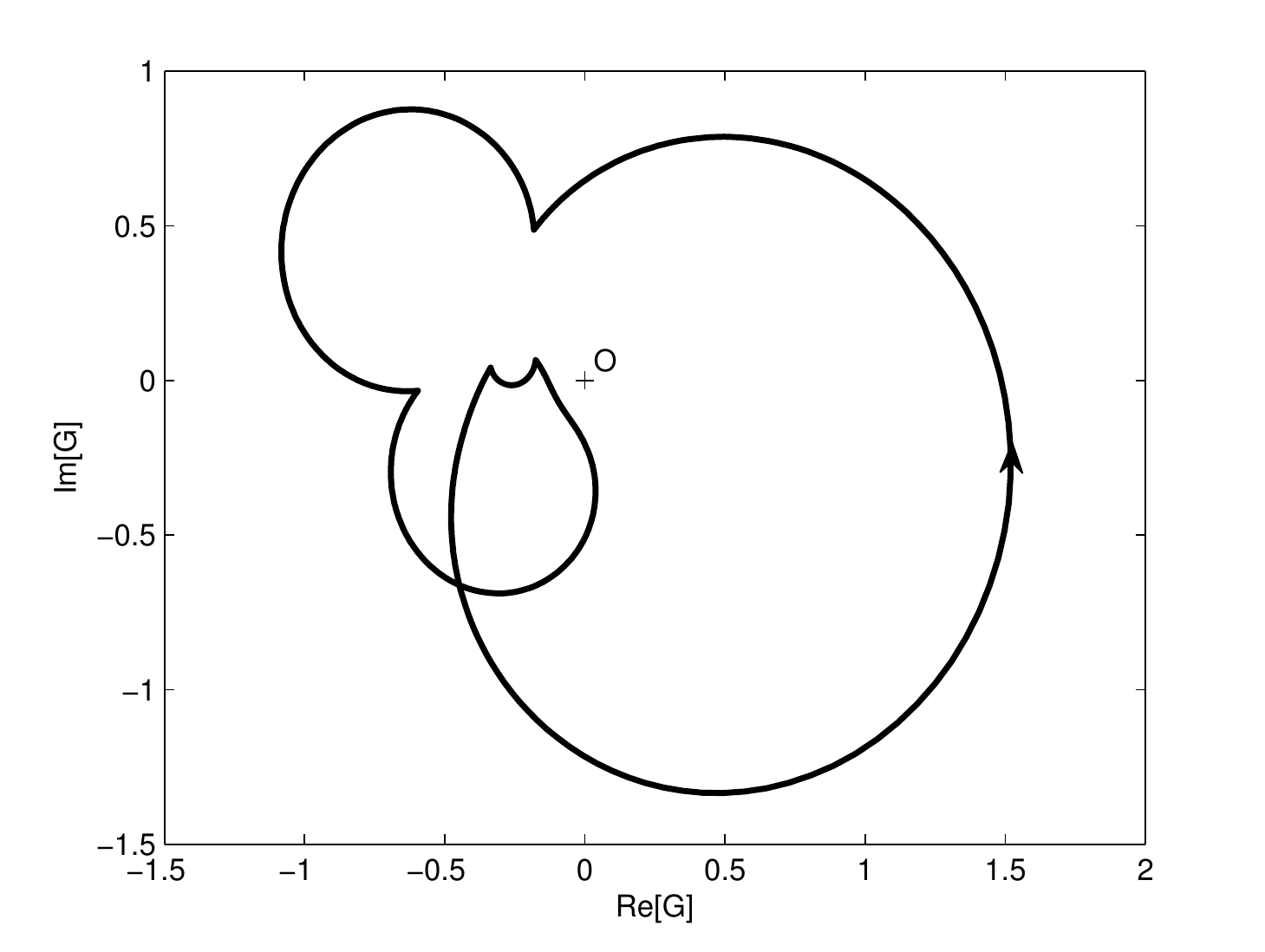}
\includegraphics[width = 8cm]{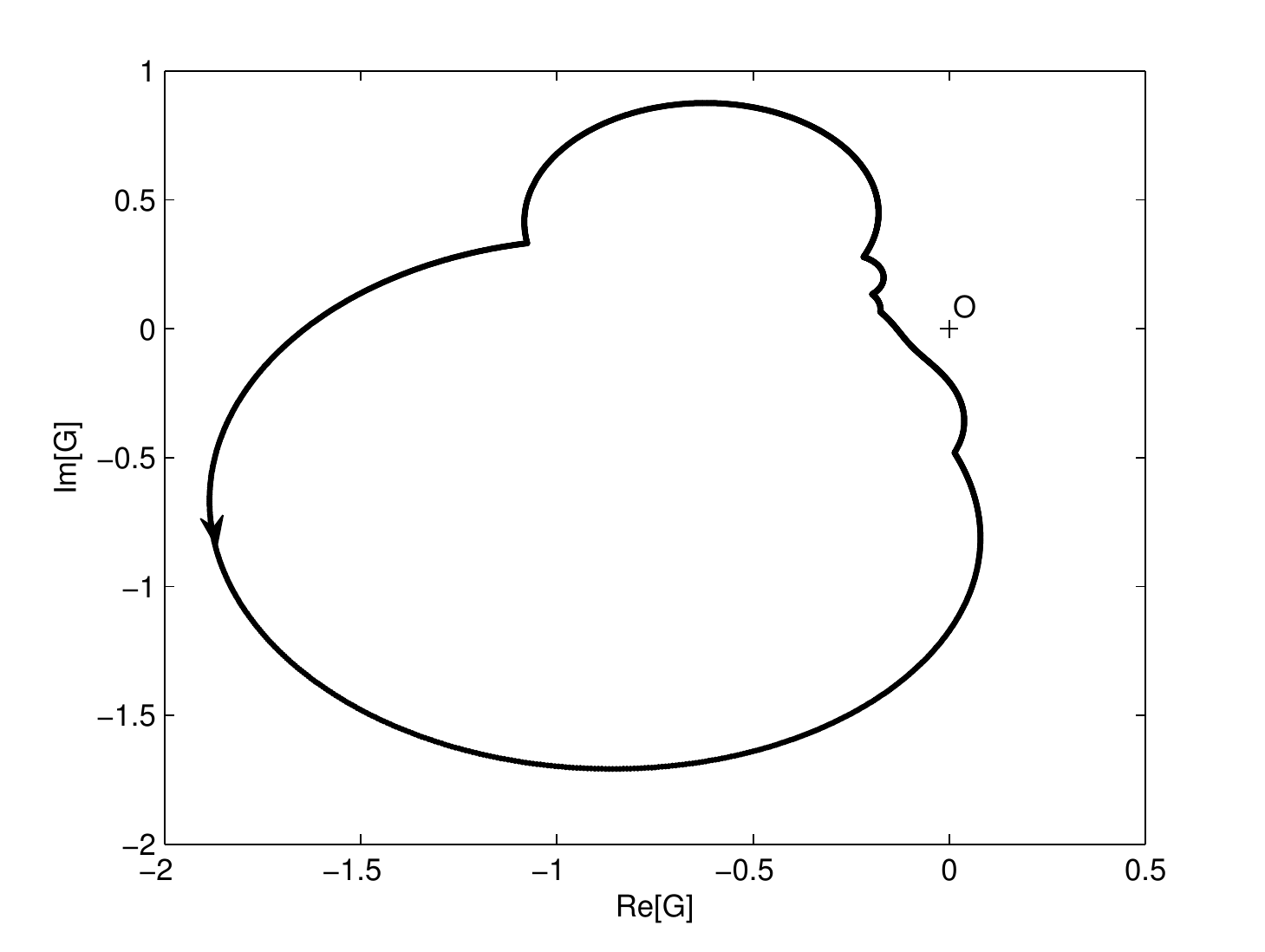}
\caption{Hodographs of $G$ on $\sigma_\alpha$ (left) and 
on $\sigma_\beta$ (right)}
\label{num_hod}
\end{figure}

The hodographs of $G_1(\hat x)$ on $\sigma_\alpha$
and $\sigma_\beta$ are shown in Fig.~\ref{num_hod}, left and right, 
respectively. The origin is marked by letter~O in both graphs. 
One can see that the hodograph for $\sigma_\alpha$
encircles the origin for a single time in the positive direction,
and the hodograph for $\sigma_\beta$ does not encircle 
the origin at all. Thus, the conditions for $G_1$ are valid. 

Indeed, a similar check can be performed for $G_2 = (F_2)^3$.

\subsection{Building the wave $u(m,n)$}

Take the value of $K$ equal to $0.5$. 
For simplicity,  take angle of incidence $\phi_{\rm in} = \pi / 4$. 
By symmetry, $x_{\rm in} = y_{\rm in}$
and we are looking for 
the solution of the equation $\hat D(x_{\rm in} , x_{\rm in}) = 0$
corresponding to the wave traveling in the positive direction with respect to $m$ 
and~$n$:
\[
x_{\rm in} = y_{\rm in} = \frac{4 - K^2 + i K \sqrt{8 - K^2}}{4} = 0.9375 + 0.3480i.
\]

For the total wave, we use an alternative form of the Sommerfeld integral
(\ref{eq212}), namely 
 (\ref{eq303}), (\ref{eq304})
with the transformant $A$ represented by (\ref{eq107}).
 We take
50000 nodes on each contour for integration in the $x$-plane. 

\begin{figure}[ht]
\centering
\includegraphics[width = 8cm]{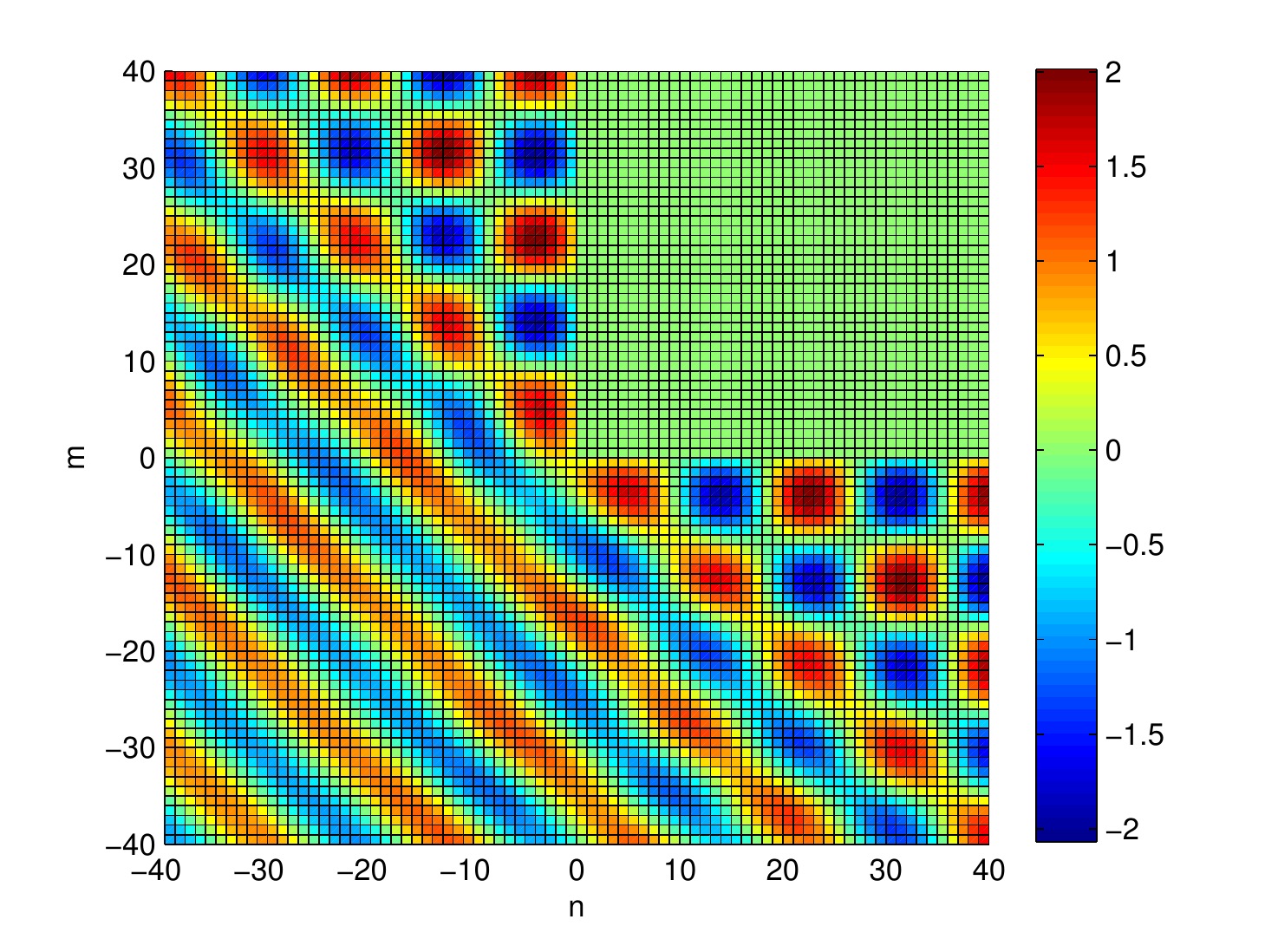}
\caption{The real part of  $u(m,n)$}
\label{fig_tot_field}
\end{figure}

The real part of the total wave is shown in Fig.~\ref{fig_tot_field}. 

The field pattern corresponds to what can be expected. The field is zero at the boundary, 
and there are visible zones  of the reflected waves. 

\begin{figure}[ht]
\centering
\includegraphics[width = 6.0cm]{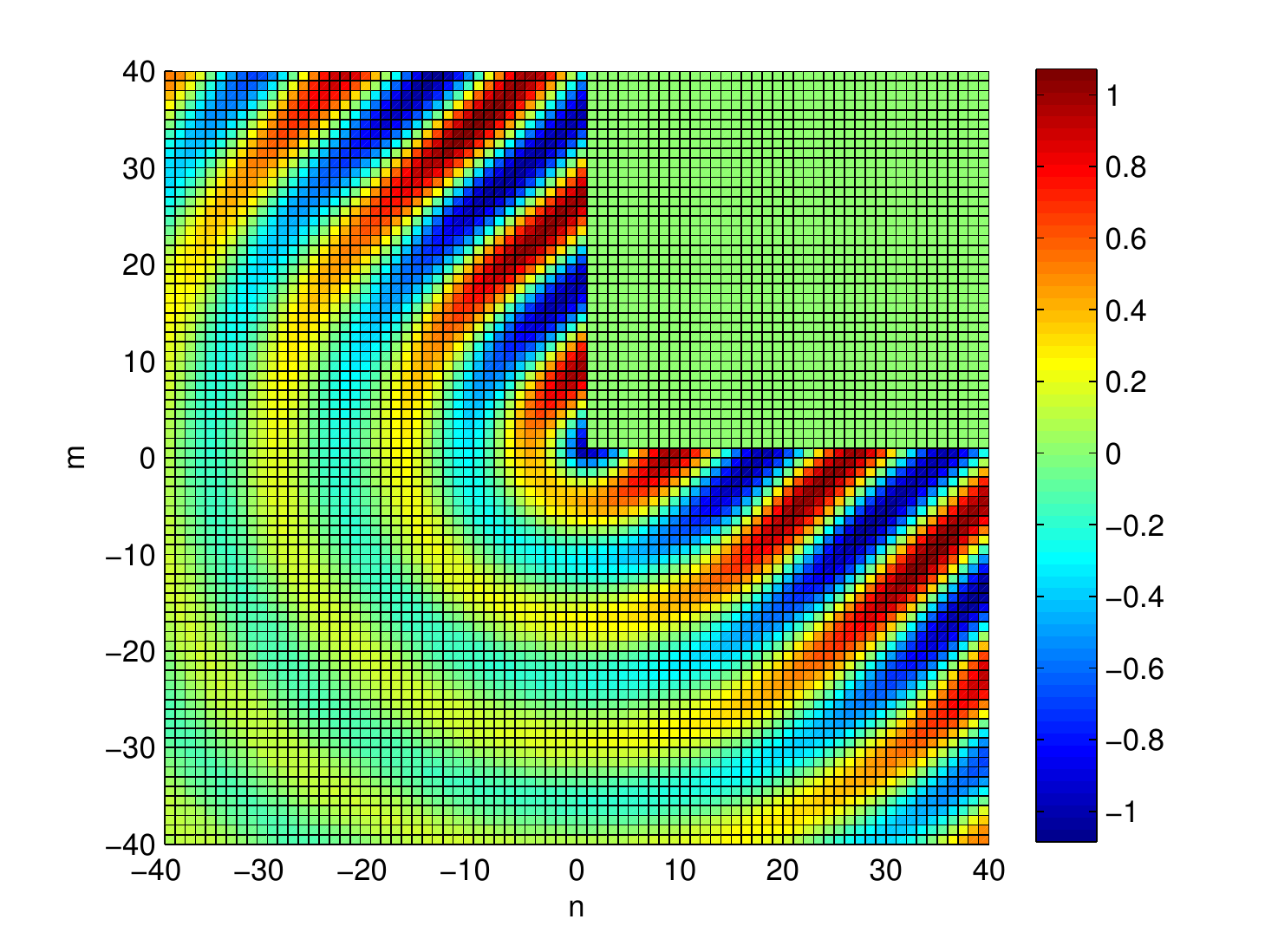}
\includegraphics[width = 6.0cm]{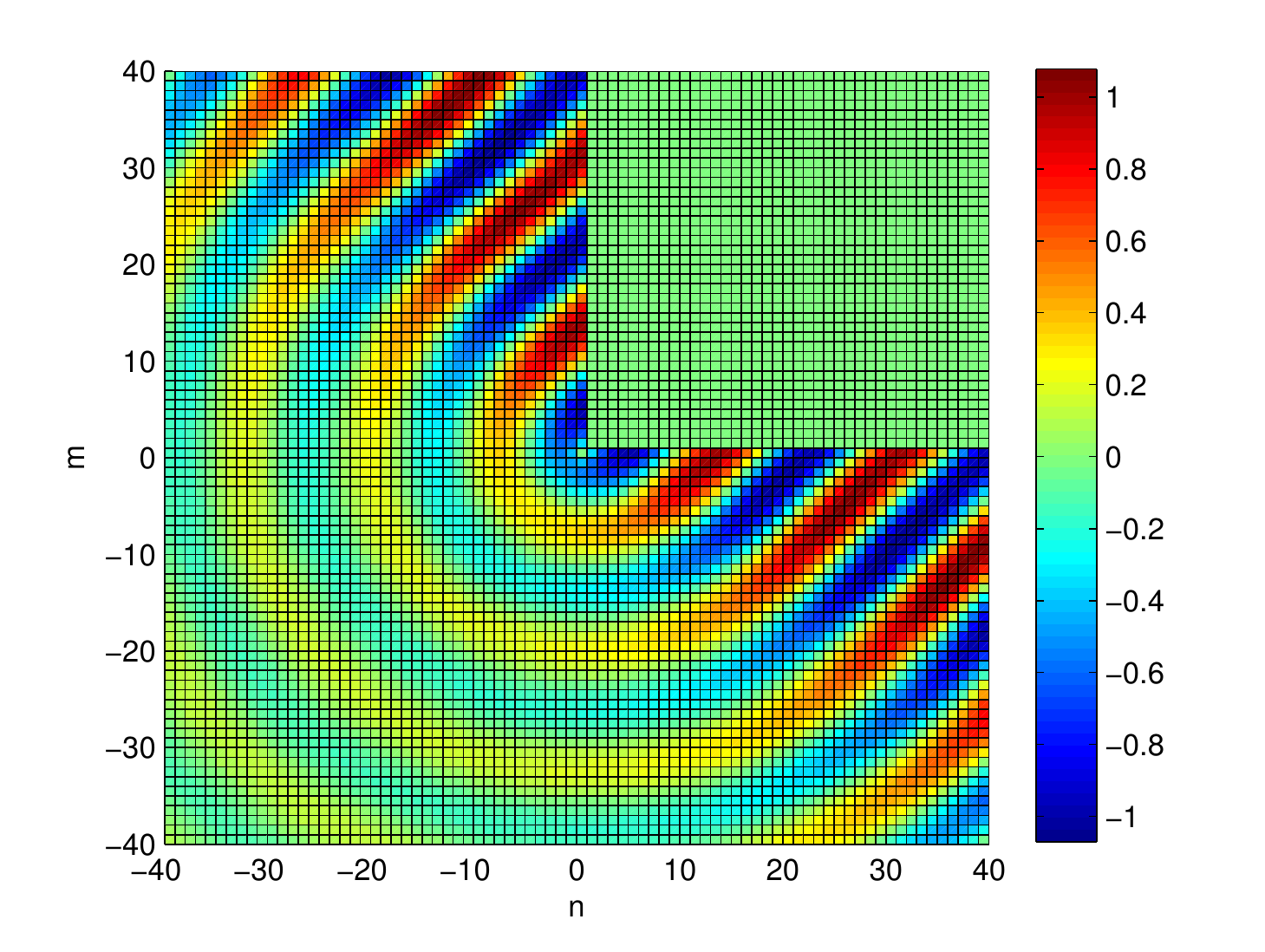}
\caption{The scattered wave $u_{\rm sc}(m,n)$. The real part (left); 
the imaginary part (right)}
\label{fig_sc_field}
\end{figure}

In Fig.~\ref{fig_sc_field} we plot the scattered wave $u_{\rm sc}$ only. 
The real part is in the left, while the imaginary part is in the right. One can see cylindrical wave scattered by the angle vertex. 

\section{Conclusion}

Let us summarize the process of solving the problem of diffraction by a Dirichlet right angle 
on a discrete plane. Note that the problem is characterized by two parameters: by the wavenumber parameter $K$ of the Helmholtz equation (\ref{eq201}) and by the incident angle $\phi_{\rm in}$
defined by (\ref{eq_phiin}). The procedure is as follows: 

\begin{enumerate}

\item 
The consideration is based on the structure of the Riemann surface~$\gR$.
This surface is described by four branch points $\eta_{1,1}$,
$\eta_{1,2}$, $\eta_{2,1}$, $\eta_{2,2}$. These branch points depend only on $K$, and they 
are found from (\ref{eq210}), (\ref{eq211}). 

\item
One should find the period $T_\beta$. This can be done by computing the 
corresponding integral in (\ref{eq005b}). The contour is shown in Fig.~\ref{fig22},  
but it is practical to 
perform the integration along
the unit circle in the negative direction. 
Function $\Upsilon(\hat x)$ is given by the last expression of (\ref{eq209b}). 
The branch of $\Upsilon(\hat x)$ is chosen in such a way that 
$y (\hat x)$ defined by (\ref{useful})
has property $|y (\tilde x)| < 1$. 
 
\item
The parameter $\hat \b$ should be found. This is the point
on $\gR$, thus it is characterized by the affix $\b$
and the branch of $\Upsilon (\hat \b)$.
   
Algorithm~1 described in Subsection~\ref{subsec_cb} can be used for this.  
According to this algorithm, first, the differential equation (\ref{eq022})
is solved numerically on the segment $\chi \in [0, T_\beta /3]$, 
and an approximation $\b'$ of the 
parameter $\b$ becomes obtained. Besides, the sheet of $\gR$ on which 
$\hat \b$ is located becomes determined. 
Second, an algebraic equation (\ref{eq019a}) is solved 
iteratively using $\hat \b'$ as the starting approximation. 

\item 
The functions $F_1 (\tilde x)$ and $F_2(\tilde x)$ are constructed by (\ref{eq018b})
and (\ref{eqF2}). Note that these functions depend on $K$ as on a parameter.

\item 
The Sommerfeld transformant of the total wave $A(\tilde x)$ is built using
(\ref{eq617})--(\ref{eq624}),
(\ref{eq602})--(\ref{eq607}),
(\ref{eq107}),
(\ref{eq018b}),
(\ref{eqF2}). 

\item
The function $u (m,n)$ is built using the Sommerfeld integral (\ref{eq212}).

\end{enumerate}


The whole consideration is held in the framework of the Sommerfeld integral. The structure of the integral may seem slightly unusual, however, as we demonstrate in 
\cite{Shanin2020}, it is a natural generalization of the Sommerfeld integral for 
angular domains known for the continuous case.

We  
build the functional field $\gK_3$ to which the Sommerfeld transfomant  belongs.  
This field is common for all incident angles. The functional field $\gK_3$ is  
represented as the basis $\Omega_{3:1}$ composed of three functions, two of which, 
$F_1$ and $F_2$ should be built. 
The construction of the basis is a non-trivial procedure. 
Then, for a particular angle of incidence 
$\phi_{\rm in}$
we find the Sommerfeld transformant 
$A(\tilde x)$. 
This task is tedious, but quite simple. The coefficients $q_j' (x)$, $g_j(x)''$ are rational functions, 
and one should find these functions obeying some known restrictions and having some 
known poles. 
This structure of solution seems to be deeply linked with the embedding procedure 
\cite{Biggs2006,Skelton2008}.

\section{Acknoledgements}
Authors are grateful to Anastasia Kisil for valuable discussions. The work is supported by the RFBR grant 19-29-06048.

\bibliography{Torus2_bibl}
\bibliographystyle{unsrt}

\section*{Appendix~1. Proof of Theorem~\ref{th_previous}}

\label{sub_why}

\vskip 6pt

\noindent
{\bf Consistency of the Sommerfeld integral}

\nobreak
To prove statement a) of the theorem, it is sufficient to show that 
\begin{equation}
\int \limits_{J_3} w_{m,n} ( x , y(\hat x)) 
\, A(\tilde x) \frac{dx}{\Upsilon (\hat x)} 
=
\int \limits_{J'_3} w_{m,n} ( x , y(\hat x)) 
\, A(\tilde x) \frac{dx}{\Upsilon (\hat x)} 
= 0
\label{eq301}
\end{equation}
for $m \le 0$ and $n \le 0$.
One can see that both $x$ and $y(\hat x)$ tend 
to~$\infty$ at the infinities of the sheets~2
and~4 of $\gR_3$. Thus, the function $w_{m,n}(x , y(\hat x))$ (see (\ref{eq206}))
does not grow as $|x| \to \infty$. According to the conditions imposed on $A$,
the integrals are equal to zero. 

\vskip 6pt
\noindent
{\bf Validity of the discrete Helmholtz equation}

\nobreak

We are starting to prove validity of statement b) of the theorem. 
Substitute the representation (\ref{eq212}) with (\ref{eq213}) for $m < 0$, or 
with (\ref{eq214}) for $n < 0$ into the equation
(\ref{eq201}). Note that the the discrete Laplace operator acts only on~$w$.
A direct check shows that (\ref{eq201}) is valid.

\vskip 6pt
\noindent
{\bf Radiation condition}

\begin{figure}[ht]
\centering
\includegraphics[width=0.8\textwidth]{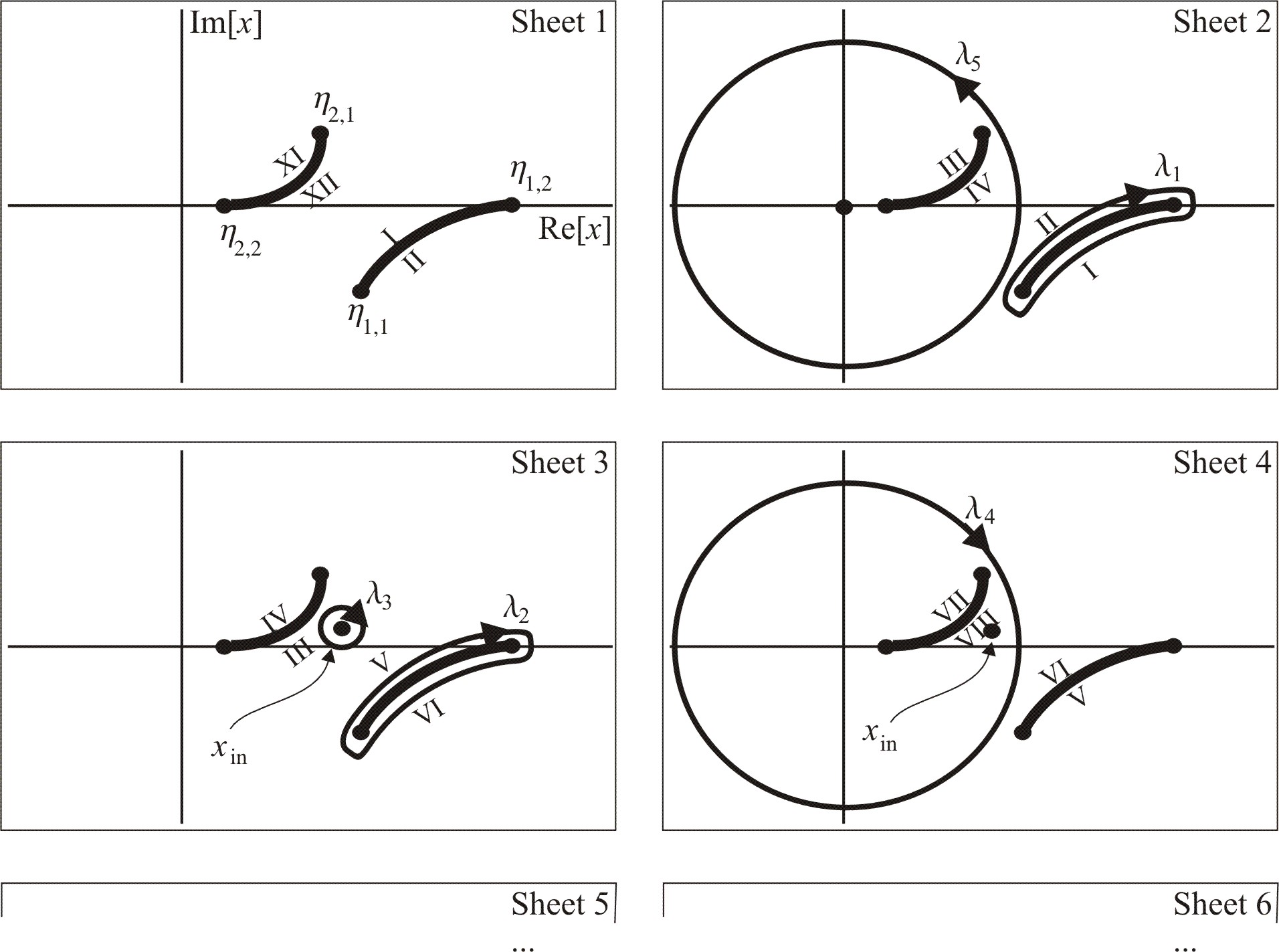}
\caption{Contours of integration $\lambda_1$, \dots , $\lambda_5$}
\label{fig24}
\end{figure}

\nobreak 
Let us demonstrate that  $u(m,n)$ obeys the radiation condition 
formulated in the form of the limiting absorption principle. 
For this, deform the contours $\Gamma_2$ and $\Gamma_3$ homotopically as follows: 
\begin{equation}
\Gamma_2 = \lambda_1 + \lambda_2 + \lambda_3,
\qquad 
\Gamma_3 = \lambda_3 + \lambda_4 + \lambda_5,
\label{eq:condef}
\end{equation}
where contours $\lambda_1, \dots , \lambda_5$ are shown 
in Fig.~\ref{fig24}. Sheets~5 and~6 are not shown. 
Contours $\lambda_1$ and $\lambda_2$ are drawn around corresponding cuts (we remind that 
the cuts are conducted along the sets of $x$ for which $|y(\hat x)| = 1$). Contours 
$\lambda_4$ and $\lambda_5$ are unit circles. Contour $\lambda_3$ encircles~$x_{\rm in}$.
Note that $|x_{\rm in}| < 1$. 

One can see that 
\begin{equation}
\int \limits_{\lambda_3} w_{m,n}(\hat x , y (\hat x)) A(\tilde x)
\frac{dx}{\Upsilon (\hat x)} = u_{\rm in}. 
\label{eq302}
\end{equation}

As the result, the following representations of the field are 
obtained: 
\begin{equation}
u(m,n) = u_{\rm in} (m,n) + \int_{\lambda_1 + \lambda_2} w_{m,n}(\hat x , y (\hat x)) A(\tilde x)
\frac{dx}{\Upsilon (\hat x)}
\qquad \mbox{for } m \le 0, 
\label{eq303}
\end{equation}
\begin{equation}
u(m,n) = u_{\rm in} (m,n) + \int_{\lambda_4 + \lambda_5} w_{m,n}(\hat x , y (\hat x)) A(\tilde x)
\frac{dx}{\Upsilon (\hat x)}
\qquad \mbox{for } n \le 0. 
\label{eq304}
\end{equation}

Consider the exponential factor $w_{m,n} = x^m y^n$ of the representation (\ref{eq303}).
For each point of the representation contours, $|y| = 1$ and 
$|x| > 1$. Since $m \le 0$, the result should decay for large negative~$m$.
Besides, 
the field should decay for constant negative $m$ and growing positive $n$
due to the oscillatory nature of factor~$w$ on the contours $\lambda_1$ and~$\lambda_2$.

Similarly, for the representation (\ref{eq304}), $|x| = 1$ and $|y| > 1$ in the 
exponential factor, thus the field should decay for large negative~$n$.
 
Thus, we obtain that the total field is a 
sum of the incident field and a decaying field. 

\vskip 6pt
\noindent
{\bf Boundary conditions}

\nobreak
Let us check the boundary condition $u= 0$ on the side $m \ge 0$, $n = 0$. 
For this, use  the representation (\ref{eq212}), (\ref{eq214}).

\begin{figure}[ht]
\centering
\includegraphics[width=0.8\textwidth]{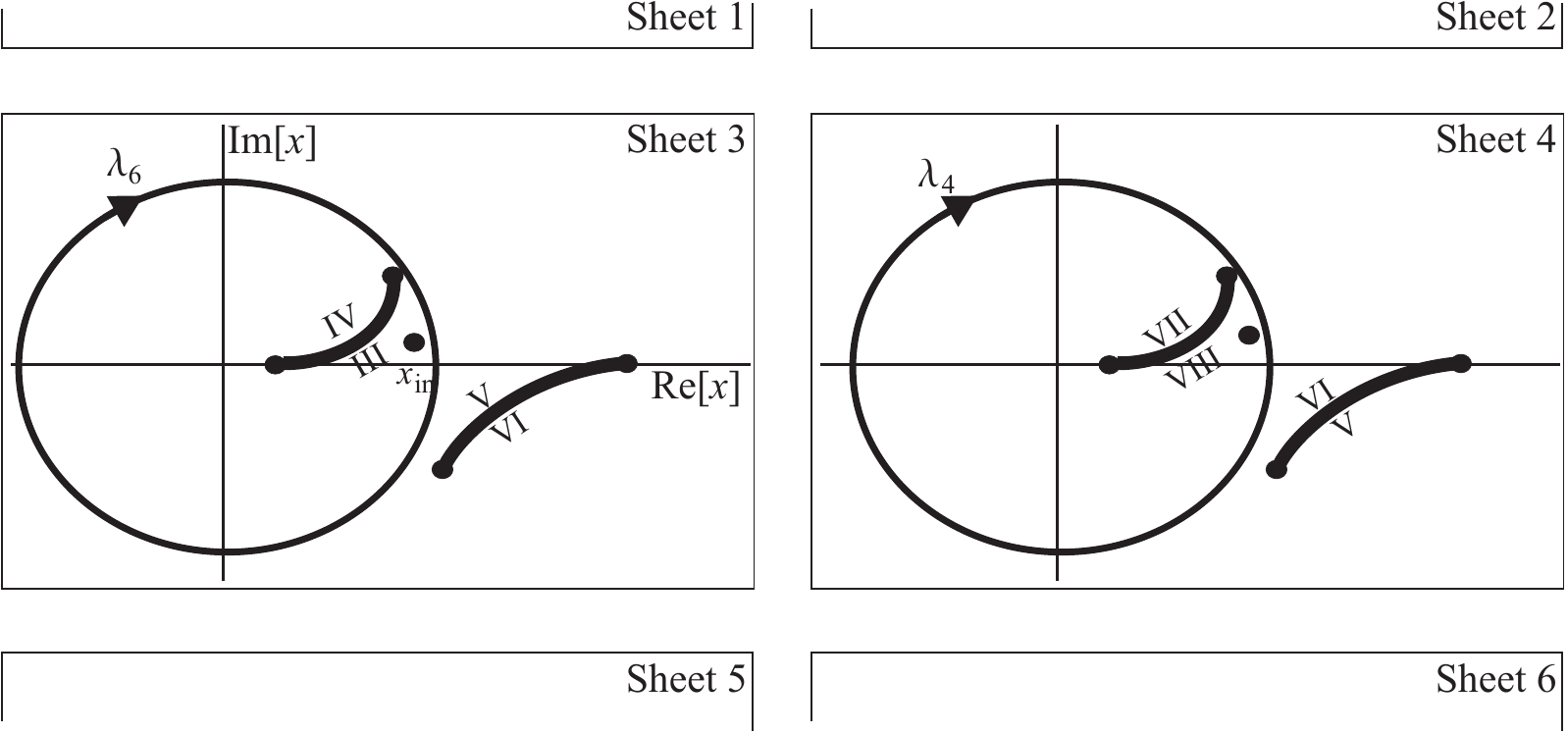}
\caption{Contours of integration $\lambda_4$ and $\lambda_6$}
\label{fig25}
\end{figure}

On the boundary $m \ge 0$, $n = 0$,
the contour of the Sommerfeld integral can be deformed into two 
unit circles drawn in sheet~3 and~4 (see Fig.~\ref{fig25}). Namely, 
\begin{equation}
u(m,0) = \int_{\lambda_4 + \lambda_6} x^m A(\tilde x) \frac{dx}{\Upsilon (\hat x)}.
\label{eq307}
\end{equation}
Due to the symmetry (\ref{eq306}), this integral is zero. 

The situation is slightly more subtle with the boundary $m = 0$, $n \ge 0$. 
Instead of $\Pi$, we need 
another symmetry $\Pi'$ of the Riemann surface~$\gR_3$. 
The symmetry $\Pi'$ is defined as 
\begin{equation}
\Pi'(\tilde X(x , 2)) = \tilde X (x^{-1} , 2)
\label{eq_Pi_pr_a}
\end{equation}
in the neighborhood of the point $x = 1$ on sheet~2, and then continued 
analytically onto the whole $\gR_3$.  
The mapping $\Pi' $ transforms $\gR_3$ into $\gR_3$ keeping its 
complex structure, but it is not a desk transformation, since it does not maintain the affix. 
Note that $y(\tilde x) = y(\Pi' (\tilde x))$. 

One can show that the function $A(\Pi' (\tilde x ))$ obeys the same functional 
problem as $A(\tilde x)$. Besides, there exists a stationary point of $\Pi'$, 
\begin{equation}
A(\Pi' (\tilde x )) = A(\tilde x). 
\label{eq_Pi_pr_b}
\end{equation}

\begin{figure}[ht]
\centering
\includegraphics[width=0.4\textwidth]{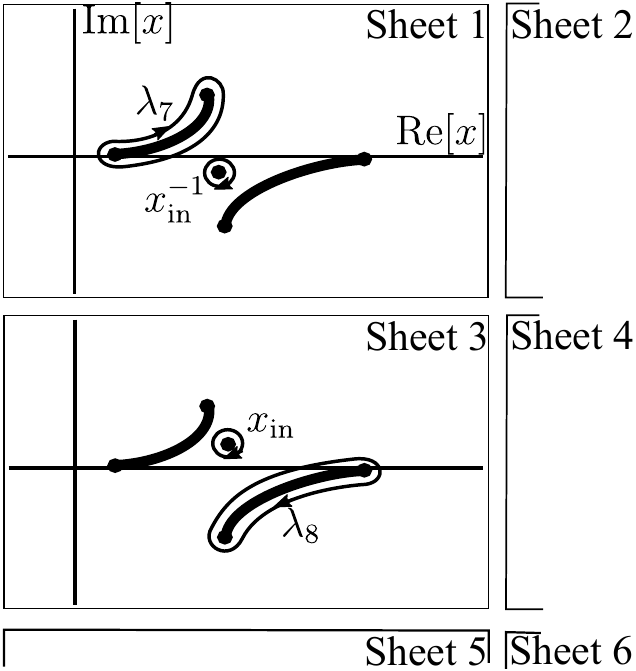}
\caption{Contours of integration $\lambda_7$ and $\lambda_8$}
\label{fig28}
\end{figure}

The contour of integration $\Gamma_2$ can be transformed into the contour shown in Fig.~\ref{fig28}
(it is composed of $\lambda_7$, $\lambda_8$, and two polar terms). 
The resulting contour is symmetrical with respect to the mapping~$\Pi'$. 
The wave factor is also symmetrical: 
\[
w_{0,n} (\tilde x, y(\tilde x)) = w_{0,n} (\Pi'(\tilde x), y(\Pi'(\tilde x))). 
\]
Thus, finally, the Sommerfeld integral on the boundary yields zero due to the symmetry. 

\section*{Appendix~2. Two propositions for algebraization of the Abel's theorem}

Here we put two more cases of the criterion of building the function~$M$.

\begin{proposition}
\label{prop_M1}

Let $a_1 , a_2, b_1, b_2$ are all distinct values not equal to infinity 
or to $\eta_{j,l}$. A function $M(\hat b_1 , \hat b_2 , \hat a_1 , \hat a_2 ; \hat x)$
exists iff
\[
\frac{ \Upsilon(\hat a_2)}{(a_2-b_1)(a_2-b_2)} 
+ \frac{\Upsilon (\hat b_1)}{(a_2 - b_1)(b_1 - b_2)} 
+ \frac{\Upsilon (\hat b_2)}{(a_2 - b_2)(b_2 - b_1)} =
\qquad \qquad
\]
\begin{equation}
\qquad \qquad 
\frac{  \Upsilon (\hat a_1)}{(a_1-b_1)(a_1-b_2)} 
+ \frac{  \Upsilon (\hat b_1)}{(a_1 - b_1)(b_1 - b_2)} 
+ \frac{  \Upsilon (\hat b_2)}{(a_1 - b_2)(b_2 - b_1)} . 
\label{eq010}
\end{equation}
\end{proposition}

%

\noindent 
{\bf Proof. }
Let us try to construct function $M(\hat b_1 , \hat b_2 , \hat a_1 , \hat a_2 ;\hat x)$ explicitly. 
First,  construct 
function $M \in \gK_1$ (possibly depending on some parameters) 
having poles only at $\hat b_1$ and~$\hat b_2$. An obvious Ansatz, up to a common constant factor,
is as follows: 
\begin{equation}
M(\hat x) = 
\frac{\Upsilon (\hat x)}{(x-b_1)(x-b_2)} + \frac{g_1}{x - b_1} + \frac{g_2}{x - b_2} + c.
\label{eq007}
\end{equation}
for arbitrary complex values $g_1$, $g_2$, $c$.
As above, $x , a_1, a_2, b_1, b_2$
are the affixes of the points 
$\hat x , \hat a_1, \hat a_2, \hat b_1, \hat b_2$, respectively.
 
For arbitrary $g_1$, $g_2$, this function has poles at four points of 
$\gR$: at $\hat b_1$, $\hat b_2$, $\Pi (\hat b_1)$, $\Pi (\hat b_2)$. Choose the values 
of $g_1$, $g_2$ such that they suppress the poles at  $\Pi(\hat b_1)$, $\Pi(\hat b_2)$. 
One can see that the appropriate function 
is as follows: 
\begin{equation}
M(\hat x) = 
\frac{\Upsilon (\hat x)}{(x-b_1)(x-b_2)} 
+ \frac{\Upsilon (\hat b_1)}{(x - b_1)(b_1 - b_2)} 
+ \frac{\Upsilon (\hat b_2)}{(x - b_2)(b_2 - b_1)} + c.
\label{eq008}
\end{equation}
 
Now let us fix the zeros.  
Choose parameter $c$ in such a way that $M(\hat a_1) = 0$: 
\begin{equation}
c = - \left[ 
\frac{\Upsilon (\hat a_1)}{(a_1-b_1)(a_1-b_2)} 
+ \frac{ \Upsilon (\hat b_1)}{(a_1 - b_1)(b_1 - b_2)} 
+ \frac{ \Upsilon (\hat b_2)}{(a_1 - b_2)(b_2 - b_1)} 
\right].
\label{eq009}
\end{equation}

Finally, the condition guaranteeing 
that $\hat a_2$ is also a zero is the equation (\ref{eq010}).
$\square$

We claim that (\ref{eq010}) is an algebraic analog of the analytic equation (\ref{eq006})
in the case of distinct $a_1, a_2 , b_1 , b_2$. 
The function $M(\hat b_1 , \hat b_2 , \hat a_1 , \hat a_2 ;\hat x)$ 
is given by (\ref{eq008}), (\ref{eq009}). 


\begin{proposition}
\label{prop_M2}
Let be $\hat b_1 = \hat b_2 (= \hat b)$, $a_{1,2} \ne b$, $a_1 \ne a_2$, and neither of the points $a_{1,2} , b$
is equal to $\eta_{j,l}$ or infinity. 
A function  $M(\hat b , \hat b , \hat a_1 , \hat a_2 ;\hat x)$ exists iff
\begin{equation}
\frac{\Upsilon (\hat a_1)}{(a_1 - b)^2}
+ 
\frac{\Upsilon (\hat b)}{(a_1 - b)^2}  
+ 
\frac{\dot \Upsilon( \hat b)}{a_1 - b} 
=
\frac{\Upsilon (\hat a_2)}{(a_2 - b)^2}
+ 
\frac{\Upsilon (\hat b)}{(a_2 - b)^2}  
+ 
\frac{\dot \Upsilon (\hat b)}{a_2 - b} .
\label{eq014}
\end{equation} 
\end{proposition}

\noindent
{\bf Proof. }
The most general Ansatz for function $M \in \gK_1$ having a pole of order~2 at $\hat b$  
is as follows (this Ansatz does not include a common constant factor): 
\begin{equation}
M(\hat x) = \frac{\Upsilon(\hat x)}{(x - b)^2}
+ 
\frac{\Upsilon (\hat b)}{(x - b)^2}  + \frac{\dot \Upsilon(\hat b)}{x - b} + c,
\label{eq012}
\end{equation}
where 
\begin{equation}
\dot \Upsilon (\hat x) \equiv \frac{d \Upsilon (\hat x)}{dx}.
\label{eq012a}
\end{equation}
The constant $c$ is chosen in such a way that $M(\hat a_1) = 0$:
\begin{equation}
c = - \left[
\frac{\Upsilon (\hat a_1)}{(a_1 - b)^2}
+ 
\frac{\Upsilon (\hat b)}{(a_1 - b)^2}  
+ 
\frac{\dot \Upsilon( \hat b)}{a_1 - b} 
\right]
\label{eq013}
\end{equation}
Finally, this function is zero at $\hat a_2$ if 
(\ref{eq014}) is valid.
The function $M$  is given by (\ref{eq012}), (\ref{eq013}). 
$\square$

\section*{Appendix~3. Proof of Theorem~\ref{th_F}}

\noindent
All three functions in the numerators and denominators of (\ref{eq017}), (\ref{eq017a}) do exist 
according to Theorem~\ref{th_Abel} and the condition (\ref{eq016}).
The order of poles and zeros can be checked directly. Thus, the
statements a) and b) of the theorem are valid by construction. 

Let us prove c) (statement d) is similar). Note that $F_1 = G_1^{1/3} (\hat x)$ is a three-valued function 
of $\hat x$ having no branch points over $\gR$, since the pole and the zero have order~3. Thus, $F_1$
has some 3-sheet Riemann surface $\gR'$ without branching over~$\gR$.
Consider the projection $P: \gR' \to \gR$.
According to Proposition~\ref{prop_cont}~b), it is sufficient to
show that $P^{-1}(\sigma_\beta)$ is three disjoint copies of $\sigma_\beta$
and that  
$P^{-1}(\sigma_\alpha)$ is a connected 3-sheet covering of $\sigma_\alpha$, 
to prove that $\gR'$ is~$\gR_3$, and thus that $F_1$ is meromorphic on~$\gR_3$.

To establish the validitity Proposition~\ref{prop_cont}~b) one should study the variation of ${\rm Arg [G_1]}$
along the contours $\sigma_\alpha$ and $\sigma_\beta$. Denote these variations by the ${\rm Var}$ symbol. 
Note that such a variation is $2\pi j$, $j \in \mathbb{Z}$, since $G_1$ is meromorphic on $\gR$ and the contours are 
closed.  
One should prove that 
\begin{equation}
\frac{{\rm Var}_{\sigma_\beta} {\rm Arg}[G_1]}{2\pi} \equiv 0 \, ({\rm mod}\, 3) 
\label{eq_var1}
\end{equation}
and
\begin{equation}
\frac{{\rm Var}_{\sigma_\alpha} {\rm Arg}[G_1]}{2\pi} \ne 0 \, ({\rm mod}\, 3) 
\label{eq_var1a}
\end{equation}
to establish the validity of Proposition~\ref{prop_cont}~b).
Moreover, the condition 
\begin{equation}
\frac{{\rm Var}_{\sigma_\alpha} {\rm Arg}[G_1]}{2\pi} \equiv 1 \, ({\rm mod}\, 3),  
\label{eq_var2}
\end{equation}
guarantees (\ref{eq_prop1}), since  
\[
{\rm Arg}[F_1] |_{\tilde x}^{\Lambda( \tilde x)} = {\rm Var}_{\sigma_{\alpha}} {\rm Arg}[F_1] = 
\frac{1}{3} {\rm Var}_{\sigma_{\alpha}} {\rm Arg}[G_1],
\]
(note that $P_{3:1}^{-1} (\sigma_{\alpha})$ connects $\eta_{2,1}$ with $\Lambda(\eta_{2,1})$).
For $G_2$, (\ref{eq_var2}) should be replaced 
with     
\begin{equation}
\frac{{\rm Var}_{\sigma_\alpha} {\rm Arg}[G_2]}{2\pi} \equiv -1 \, ({\rm mod}\, 3).  
\label{eq_var3}
\end{equation}

Let us prove (\ref{eq_var1}), (\ref{eq_var2}).  
Consider the points $\hat a$, $\hat b$, $\hat c$ obeying the condition of Theorem~\ref{th_F}. Note that
$\hat b$ and $\hat c$ are functions of $\hat a$:
\begin{equation}
\hat b = \hat b( \hat a) = \psi ( \chi(\hat a) + T_\beta / 3 ), 
\qquad
\hat c = \hat c( \hat a) = \psi ( \chi(\hat a) + 2 T_\beta / 3 ). 
\label{eq_bc}
\end{equation}

\begin{figure}[ht]
\centerline{\epsfig{file=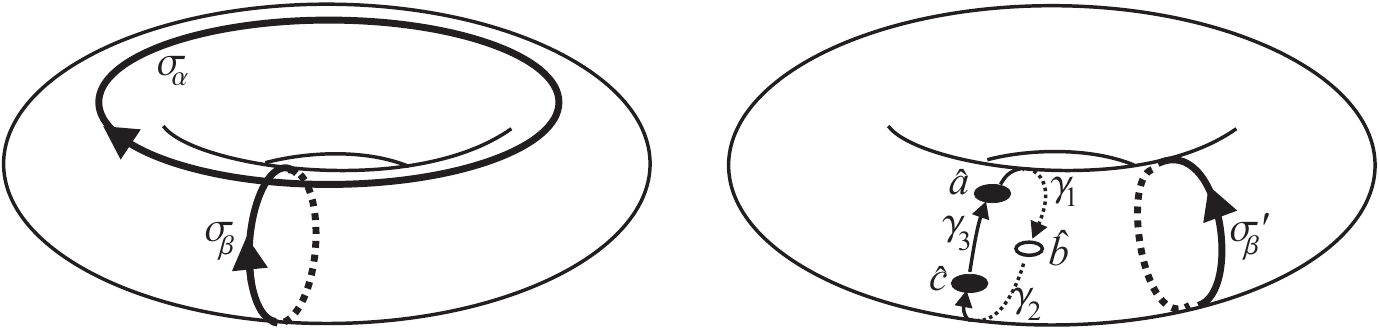, width = 12cm}}
\caption{Contours $\sigma_\alpha$ and $\sigma_\beta$ on $\gR$ (left).
Points $\hat a$, $\hat b$, $\hat c$ on the torus $\gR$ (right)}
\label{fig01}
\end{figure}

The image $\chi(\sigma_\beta)$ is a continuous path connecting the points $0$ and $T_{\beta}$.
Deform $\sigma_\beta$ into $\sigma'_\beta$ such that $\chi(\sigma'_\beta)$ is a {\em straight segment\/}
connecting these points.
During this deformation, the variation 
${\rm Var}_{\sigma_\beta} {\rm Arg}[G_1]$ can change only by $6\pi j$, since the contour can 
cross the poles or zeros of order~3.
Deform $\sigma_\alpha$ the same way, i.~e.\ such that  $\chi(\sigma'_\alpha)$ is a straight segment connecting 
$0$ and~$T_\alpha$. Take $\gamma_1$, $\gamma_2$, $\gamma_3$ such that 
$\chi(\gamma_1 + \gamma_2 + \gamma_3)$ is a straight line connecting $\chi(\hat a)$ with 
$\chi(\hat a) + T_\beta$.  Note that $\chi(\gamma_j)$ become parallel $\chi(\sigma_\beta)$, thus generally 
$\gamma_j$ do not cross~$\sigma'_\beta$ (see Fig.~\ref{fig01}). 

Define a function 
\begin{equation}
M'(\hat a ; \hat x) = \frac{ M(\hat a , \hat a , \hat b(\hat a) , \hat c(\hat a) ; \hat x)}{
M(\hat a , \hat a , \hat b(\hat a) , \hat c(\hat a) ; \infty)} .
\label{eq_bc1}
\end{equation}   
One can see that the function is normalized for definiteness; the infinity point is taken on sheet~1. 
According to (\ref{eq017}), one can see that (up to a constant factor)
\begin{equation}
G_1 = M'(\hat a ; \hat x) / M' (\hat b ; \hat x). 
\label{eq_bc2}
\end{equation}

By construction (see the explicit formulae from Proposition~\ref{prop_M2}), 
function $M'(\hat a ; \hat x)$ depends analytically on $\hat a$. Let the point $\hat a'$
move from $\hat a$ to $\hat b$ along~$\gamma_1$. Note that $\gamma_1$ does not cross 
$\sigma_\beta$, thus, as $\hat a'$ travels along $\gamma_1$, the argument variation 
of $M'(\hat a' ; \hat x)$ along $\sigma'_\beta$ does not change. Thus, 
\begin{equation}
{\rm Var}_{\sigma'_\beta} [{\rm Arg}[G_1]] = 0 .
\label{eq_bc3}
\end{equation}

Now let us prove (\ref{eq_var2}). As $\hat a'$ travels along $\gamma_1$ from $\hat a$
to $\hat b$, $\sigma'_\alpha$ is crossed either by a double pole or by a simple zero. 
A careful study of the orientation of the contours states that 
in the first case ${\rm Var}_{\sigma'_\alpha}[G_1]$ changes by $4\pi$, and in the second case 
by $-2\pi$. In both cases (\ref{eq_var2}) is valid. $\square$ 


\section*{Appendix~4. Uniqueness of the solution of the diffraction problem} 

Let there be two different scattered waves, $u_{\rm sc}(m,n)$ and $u_{\rm sc}'(m,n)$, both obeying the 
conditions of the diffraction problem. Consider the difference 
\[
v(m,n) = u_{\rm sc}(m,n) - u_{\rm sc}'(m,n). 
\] 
Let us show that it is equal to zero. Function $v$ obeys equation (\ref{eq201}), radiation condition and the 
homogeneous Dirichlet boundary conditions. Moreover, it belongs to~$l_2 (\mathbb{Z}^2)$. 

Introduce a single index $\mu$ numbering all non-boundary nodes of the domain shown in Fig.~\ref{fig12b} in a reasonable order. 
Consider the function
\[
v_\mu = v(m (\mu) , n(\mu)). 
\]
The equation (\ref{eq201}) can be rewritten in the form 
\begin{equation}
\sum_{\mu'} C_{\mu , \mu'} v_{\mu'} + K^2 v_\mu = 0,
\label{eqA401}
\end{equation}
where
\[
C_{\mu , \mu'} =   
\left \{ \begin{array}{ll}
1 & \mu \mbox{ and } \mu' \mbox{ are neighboring nodes},\\
-4 & \mu = \mu' ,\\
0  & \mbox{otherwise}
\end{array}\right.  
\]
Note that $C_{\mu, \mu'} = C_{\mu', \mu}$, and $C_{\mu, \mu'}$ is a real (infinite) matrix. 
Take the complex conjugation of (\ref{eqA401}):
\begin{equation}
\sum_{\mu'} C_{\mu , \mu'} \bar v_{\mu'} + \bar K^2 \bar v_\mu = 0,
\label{eqA402}
\end{equation}
Multiply (\ref{eqA401}) by $\bar v_\mu$. Perform summation over all~$\mu$.
Multiply (\ref{eqA402}) by $ v_\mu$ and also perform a summation. Subtract the second sum from the first one. 
The result is 
\[
(K^2 - \bar K^2) \sum_{\mu} \bar v_\mu v_\mu = 0 .
\]
Indeed, this yields $v_\mu \equiv 0$.

\end{document}